\documentclass[journal,a4paper,twocolumn]{IEEEtran}

\usepackage{graphicx}
\usepackage{subfigure}
\usepackage{cite}
\usepackage{url}
\PassOptionsToPackage{hyphens}{url}
\usepackage[colorlinks]{hyperref}

\hypersetup{
    linkcolor=blue,
    filecolor=magenta,      
    urlcolor=teal,
    citecolor=magenta
    }
\usepackage{breakurl}
\usepackage{algorithm}
\usepackage{algorithmic}

\usepackage{amsfonts}
\usepackage{amsmath}
 
\usepackage{amsthm}
\theoremstyle{definition}
\newtheorem{example}{Example}
\newtheorem{result}{Result}

\newcommand{\myresultbox}[1]{\begin{center}
\noindent
\parbox{0.85\columnwidth}{%
\begin{result}
\textit{#1}
\end{result}
}\end{center}}

\newcommand{\mypropositionbox}[1]{\begin{center}
\noindent
\fbox{%
\parbox{0.85\columnwidth}{%
\textbf{Proposition:}
\textit{#1}
}
}\end{center}}

\usepackage{enumitem}
\usepackage{bm} 
\usepackage{algorithm}
\usepackage{algorithmic}
\usepackage[]{vaucanson-g}

\DeclareMathOperator{\Tr}{tr}

\DeclareMathOperator*{\argmin}{\arg\!\min}

\begin{document}


\title{Mathematical Modeling of Epidemic Diseases; A Case Study of the COVID-19 Coronavirus}
\author{Reza~Sameni$^\textnormal{*}$\\
Version 4, Dec 30th 2020%
\thanks{$^\textnormal{*}$R. Sameni is an Associate Professor of Biomedical Engineering at the Department of Biomedical Informatics, Emory University School of Medicine, 101 Woodruff Circle, Atlanta, GA 30322, US  (e-mail: \url{rsameni@dbmi.emory.edu}).}%
}
\markboth{DRAFT VERSION, SUBJECT TO MODIFICATION. FOLLOW THE UPDATES FROM: \MakeLowercase{\url{https://arxiv.org/abs/2003.11371}}}{DRAFT VERSION, SUBJECT TO MODIFICATION. FOLLOW THE UPDATES FROM: \MakeLowercase{\url{https://arxiv.org/abs/2003.11371}}}
\maketitle
\begin{abstract}
In this research, we study the propagation patterns of epidemic diseases such as the COVID-19 coronavirus, from a mathematical modeling perspective. 
The study is based on an extensions of the well-known \textit{susceptible-infected-recovered (SIR)} family of compartmental models. It is shown how social measures such as distancing, regional lockdowns, quarantine and global public health vigilance, influence the model parameters, which can eventually change the mortality rates and active contaminated cases over time, in the real world. As with all mathematical models, the predictive ability of the model is limited by the accuracy of the available data and to the so-called \textit{level of abstraction} used for modeling the problem. In order to provide the broader audience of researchers a better understanding of spreading patterns of epidemic diseases, a short introduction on biological systems modeling is also presented and the Matlab source codes for the simulations are provided online.
\end{abstract}
\IEEEpeerreviewmaketitle
\section{Introduction
\label{sec:introduction}}
Since the outbreak of the COVID-19 coronavirus in early 2020, the virus has affected most countries and taken the lives of several thousands of people worldwide. By March 2020, the World Health Organization (WHO) declared the situation a pandemic, the first of its kind in our generation. To date, many countries and regions have been locked-down and applied strict social distancing measures to stop the virus propagation. From a strategic and healthcare management perspective, the propagation pattern of the disease and the prediction of its spread over time is of great importance, to save lives and to minimize the social and economic consequences of the disease. Within the scientific community, the problem of interest has been studied in various communities including mathematical epidemiology \cite{diekmann2012mathematical,brauer2012mathematical}, biological systems modeling \cite{haefner2005,deVries2006}, signal processing \cite{yang2014comparison} and control engineering \cite{azar2020control}.

The problem of pandemic modeling has significant practical importance for governments and decision-makers. Non-pharmaceutical interventions (NPIs) refer to actions and policies adopted by individuals, authorities or governments that help slowing down the spread of epidemic diseases. NPIs are among the best ways of controlling pandemic diseases when vaccines or medications are not yet available\footnote{See Centers for Disease Control and Prevention guidelines on NPIs: \url{ https://www.cdc.gov/nonpharmaceutical-interventions/}.}. During the COVID-19 pandemic, several attempts have been made to categorize and quantify the various NPIs of different regions and nations. The quantification of the NPI is believed to be helpful for comparing the effectiveness of regional policies in containing the pandemic spread. By using machine learning techniques, the quantified NPI can be used to forecast the future trends of the pandemic and to simulate \textit{``what if scenarios''} for the better management of human and medical resources, and to eventually prescribe appropriate NPI for controlling the pandemic \cite{miikkulainen2020prediction}. The Oxford COVID-19 Government Response Tracker (OxCGRT) is one of the NPI tracking projects, which were launched and regularly updated during the COVID-19 pandemic \cite{OxCGRT2020}. Most recently, this project has been used in the machine learning community to launch data challenges for NPI-based prediction and prescription plans \cite{XPRIZEPandemicResponse2020}. 

In this study, epidemic outbreaks are studied from an interdisciplinary perspective, by using an extension of the \textit{susceptible-exposed-infected-recovered (SEIR)} model \cite{brauer2012mathematical}, which is a mathematical compartmental model based on the average behavior of a population under study. The objective is to provide researchers a better understanding of the significance of mathematical modeling for epidemic diseases. It is shown by simulation, how social measures such as distancing, regional lockdowns and public health vigilance, can influence the model parameters, which in turns change the mortality rates and active contaminated cases over time. 

It should be highlighted that mathematical models applied to real-world systems (social, biological, economical, etc.) are only valid under their assumptions and hypothesis. Therefore, this research--- and similar ones--- that address epidemic patterns, do not convey direct clinical information and dangers for the public, but should rather be used by healthcare strategists for better planning and decision making. 
Hence, the study of this work is only recommended for researchers familiar with the strength points and limitations of mathematical modeling of biological systems. The Matlab codes required for reproducing the results of this research are also online available in the Git repository of the project \cite{EpidemicModelingCodes}.

In Section \ref{sec:background}, a brief introduction to mathematical modeling of biological systems is presented, to highlight the scope of the present study and to open perspectives for the interested researchers, who may be less familiar with the context. The proposed model for the outspread of the coronavirus is presented in Section \ref{sec:themodel}. The article is concluded with some general remarks and future perspectives.

\section{An introduction to mathematical epidemiology and compartmental modeling \label{sec:background}}

\subsection{Mathematical modeling}
A model is an entity that resembles a system or object in certain aspects, but is easier to work with as compared to the original system. Models are used for the 1) identification and better understanding of systems, 2) simulation of a system's behavior, 3) prediction of its future behavior, and ultimately 4) system control. Apparently, from item 1 to 4, the problem becomes more difficult and although the ultimate objective is to harness or control a system, this objective is not necessarily achievable. While modeling is the first and most important step in this path, it is highly challenging and nontrivial. The various issues that one faces in this regard, include:
\begin{itemize}[leftmargin=*]
\item Models are \textit{not unique} and different models can co-exist for a single system.
\item A model is only \textit{a slice of reality} and all models have a \textit{scope}, outside of which, they are invalid.
\item Modeling can be done in different \textit{levels of abstraction}, which corresponds to the level of simplification and the specific aspects of the system that are considered by the model. 
\end{itemize}
\begin{example}
The response of global stock markets with numerous economic, political, industrial, social and psychological factors, to a high impact news can in cases be modeled with a second-order differential equation, with a step-like over-damped behavior that reaches its steady state after a while. Or in medicine, the response of the human body--- with more that thirty-seven trillion cells--- to medication can in many cases be ``resonably'' modeled with a first order differential equation.
\end{example}

While various types of models are used for biological systems, we are commonly interested in mathematical models \cite{ottesen2004}, as they permit the prediction and possible control of biological systems. In choosing among different available models, the widely accepted principle is the model \textit{parsimony}, which simply means that \textit{``a model should be as simple as possible and as complex as necessary!''}. The model parsimony, is also an important factor for estimating the unknown model parameters using real data. A more accurate model with fewer number of parameters is evidently preferred over a less accurate and more complex model. But how should one select between a more accurate complex model and a less accurate simpler one? Measures such as the Akaike information criterion (AIC), the Bayesian information criterion (BIC) and the minimum description length (MDL), address the balance between the number of observations and the model unknown parameters to select between competing models with variable number of parameters and different levels of accuracy \cite{burnham2004multimodel,de2005modified}. Finally, the physical interpretability of the model parameters and the ability to estimate the parameters such that the model matches real-world data, is what makes the whole modeling framework meaningful.

\subsection{From stochastic infection propagation models to ordinary differential (difference) equation modeling\label{sec:stochasticode}}
The outbreak of a contagious disease in a large population is a stochastic event. Starting from a single infected individual, the infection is transmitted to others in a stochastic manner, either by direct contact, proximity, or environmental traces (infected objects left over in the environment). The new infected generation in turns transmits the infection (again probabilistically) to the healthy individuals that they meet or encounter. During the primary stages of an epidemic outbreak, healthy-infected individual encounters are statistically independent. As a result, the chance of multiple infected people meeting a single healthy individual is probabilistically low. Therefore, assuming that each infected individual contaminates $\mathcal{R}_0$ new people on average (known as the \textit{reproduction number}, more rigorously defined in Section \ref{sec:R0}), if $\mathcal{R}_0 > 1$ the disease spreads exponentially from one time step to another (for example on a daily basis). However, in a finite population, the exponential growth can not continue for ever. Depending on the population size and contact patterns, the probability of infected people encountering independent healthy individuals decreases. Therefore, after the initial outbreak that exponentially spreads among the population, the infected population tend to encounter each other and repeated healthy ones (the healthy individuals already contacted by another infected person). Hence, the stochastic model of infection propagation, somehow saturates\footnote{Another interpretation is that after a while, it becomes more and more difficult for the average infected person to meet $\mathcal{R}_0$ non-infected individuals, and therefore the reproduction number drops and the number of new infections decreases exponentially.}. The probabilistic models used for modeling such epidemic spread are commonly based on the \textit{branching process} and a \textit{Poisson distributions} for the probability of contact between infectious and healthy subjects. The stochastic perspective to epidemic modeling has been extensively studied in the literature \cite{britton2010stochastic,pellis2012reproduction,brauer2012mathematical,miller2019distribution}. Herein, we adopt a more heuristic approach for model formation, which is less rigorous, but is equally accurate in large populations (refer to the above references for the justification). 

Suppose that $x(t)$ denotes the number of infected individuals of a population at time $t$. Next, assuming that the chance of infection increases with the number of infected individuals, we assume that the variations in the population of the number of infected between time $t$ and $t + \Delta$ (over relatively small intervals $\Delta$) is proportional to the number of infected individuals, i.e.,
\begin{equation}
    \frac{d x(t)}{dt} \approx
    \frac{x(t + \Delta) - x(t)}{\Delta} = \phi(t) x(t)
    \label{eq:expo}
\end{equation}
Let us name $\phi(t)$ the \textit{reproduction function}, which models how the infected population evolves over time. This function accounts for the expectation of various probabilistic factors, such as the rate of infection transmission, population density and contact patterns. Note that although the $x(t)$ on the right hand side of (\ref{eq:expo}) could have been unified in $\phi(t)$, the above form has the advantages that $\phi(t)$ can be interpreted as the exponential rate, with inverse time units.

Denoting the $k$th generation of the infection spread by $x_k \stackrel{\Delta}{=} x(k \Delta)$, (\ref{eq:expo}) can be discretized as follows:
\begin{equation*}
    x_{k+1} = [1 + \Delta\phi(k\Delta)]x_k
\end{equation*}
Now, defining the reproduction number $r_k \stackrel{\Delta}{=}[1 + \Delta\phi(k\Delta)]$, it is evident that the population at the discretized time index $k$ can be recursively found from the initial condition $x_0$:
\begin{equation}
    x_{k} = (r_{k-1}r_{k-2}\cdots r_{0}) x_0
\end{equation}
Apparently, if for all $k$, $r_{k} < 1$ (or equivalently $\phi(t) < 0$), the infection would decay to zero; otherwise if $r_{k} > 1$ (or $\phi(t) > 0$) it spreads. In the simplest case, for which the reproduction function is a constant $\phi(t) = \lambda$, we have a constant reproduction number $\mathcal{R}_0 = 1 + \lambda\Delta$, resulting in an exponential growth/decay:
\begin{equation}
    x_k = x_0\mathcal{R}_0^k,
    \label{eq:exposolutiondiscrete}
\end{equation}
or in the continuous case: 
\begin{equation}
    x(t) = x(0) e^{\lambda t}
    \label{eq:exposolutioncont}
\end{equation}
The equivalence of the discrete and continuous solutions is evident up to the first order approximation of the derivative, as assumed in (\ref{eq:expo}). 

More generally, the reproduction function $\phi(t)$ (or $r_k$ in the discrete case) can be a time-varying function of factors such as the total susceptible population, the population of the exposed individuals (carriers of the disease but without symptoms), contact patterns, and countermeasures such as social distancing and lockdowns. As shown in the sequel, the notion of reproduction function (number) and its impact on epidemic outbreak generalizes to eigen-analysis of vector-valued dynamic epidemic models (when the population is divided into multiple groups of individuals known compartments), enabling the stability analysis of such models.

As a reminder for later use, when $\lambda < 0$, the exponential law in (\ref{eq:exposolutioncont}) implies that the population of the infected cases drops 63\%, 86\%, 95\%, 98\%, 99\%, and 99.75\% from its initial value, after $\lambda^{-1}$, $2\lambda^{-1}$, $3\lambda^{-1}$, $4\lambda^{-1}$, $5\lambda^{-1}$, and $6\lambda^{-1}$ time units, respectively. The latter indicates for example that after $6\lambda^{-1}$ time units, only 25 cases out of 10,000 would still be infected. This property is later used to estimate the model parameters from clinical experimental results. It is good to note that although the exponential law for infection spread is the most common assumption, depending on the application, more accurate non-exponential models have also been considered \cite{feng2007epidemiological}. 



\subsection{Compartmental modeling}
Differential (difference) equations arise in many modeling problems. The major application of these equations is when the rate of change of a variable is related to other variables, as it is so in most physical and biological systems. Many powerful mathematical tools exist for the analysis and (numerical) solution of models based on differential equations. Despite their vast applications, differential equations are difficult to conceive and interpret without visualization. In this context, compartmental models are used as a visual means of representing differential equations of dynamic systems. A compartment is an abstract entity representing the quantity of interest (volume, number, density, etc.). Depending on the level of abstraction, each of the variables of interest (equivalent to system states in dynamic systems) are represented by a single compartment, conceptually represented by a box. Each compartment is assumed to be internally \textit{homogeneous}, which implies that all entities assumed inside the compartment are indistinguishable. For example, depending on the model complexity selected for modeling a certain epidemic disease, men and women at risk can be assumed to conform a single compartment, or may alternatively be considered as different compartments. A similar partitioning may be considered for different age groups, ethnicities, countries, etc., at a cost of a more complex (less parsimonious) model with additional states and parameters to be identified. Apparently, the available real-world data may be insufficient for the parameter identification of a more detailed (complex) model.

The compartments interact with one another through a set of rate equations, visually represented by arrows between the compartments. Therefore, compartmental models can be converted to a set of first order linear or nonlinear equations (and vice versa), by writing the net flow into a compartment. Compartmental modeling is also known as \textit{mass transport} \cite{Rideout1991}, or \textit{mass action} \cite{Ingalls2012}, in other contexts. More technically, a compartmental model is a weighted directed graph representation of a dynamic system. Each compartment corresponds to a node of the graph and the linking arrows are the graph edges. From this perspective, for an $n$ compartment system, the compartment variables can be considered as state variables denoted in vector form as $\bm{x}(t) = [x_1(t), \ldots, x_n(t)]^T$. The compartmental model provides a graphical representation of the state-space model:
\begin{equation}
    \begin{array}{rl}
         \displaystyle\dot{\bm{x}}(t) &= \mathbf{f}(\bm{x}(t), \bm{w}(t) ; \bm{\theta}(t),t)\\
         \bm{y}(t) &= \mathbf{g}(\bm{x}(t); \bm{\theta}(t), t) + \bm{v}(t) 
    \end{array}
\label{eq:statespace}    
\end{equation}
where $\mathbf{f}(\cdot)$ is the state dynamics function corresponding to the compartmental model graph (which can be possibly time-variant and nonlinear), $\bm{w}(t)=[w_1(t), \ldots, w_l(t)]^T$ represents deterministic or stochastic external system inputs, $\bm{y}(t)=[y_1(t), \ldots, y_m(t)]^T$ is the vector of observable model variables considered as outputs (the measurements), $\mathbf{g}(\cdot)$ if the function that maps the state variables to the observations (measurements), $\bm{v}(t)=[v_1(t), \ldots, v_m(t)]^T$ is the vector of measurement inaccuracies, considered as additive noise and $\bm{\theta}(t)=[\theta_1(t), \ldots, \theta_p(t)]^T$ is a vector of model parameters to be set or identified. Researchers familiar with estimation theory, have already guessed that the state-space form of (\ref{eq:statespace}), implies that one may eventually be able to estimate and predict the compartment variables from noisy measurements, using state-space estimation techniques, such as the Kalman or extended Kalman filter \cite{GrewalAndrews01}.

With this background, the basic steps of compartmental modeling are:
\begin{enumerate}[leftmargin=*]
\item Identifying the quantities of interest as distinct compartments and selecting a variable for each quantity as a function of time. These variables are the \textit{state variables} of the resulting \textit{state-space equations}. 
\item Linking the compartments with arrows indicating the \textit{rate} of quantity flow from each compartment to another (visually denoted over the arrows connecting the compartments).
\item Writing the corresponding first-order (linear or nonlinear) differential equations for each of the state variables of the model. In writing the equations from the graph representation, the edge weights multiplied by the state variable of their start node are added to (subtracted from) the rate change equation of the end node (start node). External inputs can be considered to be originated from an external node with value 1.
\item Setting initial conditions and solving the system of equations (either analytically or numerically), which is in the form of a first-order state-space model.
\end{enumerate}
A compartmental model is \textit{linear} (\textit{nonlinear}), when its rate flow factors are independent (dependent) of the state variables. A compartmental model is \textit{time-invariant} (\textit{time-variant}), when its rate flow factors are independent (dependent) of time. Compartmental models may be \textit{open} or \textit{closed}. In closed systems, the quantities are only passed between the compartments, while in open systems the quantities may flow into or out of the whole system. In a closed compartmental model, the sum of all the differential equations of the system is zero (for all $t$).

\begin{example}
A three compartment model corresponding to the following set of equations is shown in Fig.~\ref{fig:compartmentalexample}.
\begin{equation}
 \begin{array}{l}
 \displaystyle\frac{dx(t)}{dt} = u - \gamma x(t)^2 - \alpha x(t)\\
 \displaystyle\frac{dy(t)}{dt} = \alpha x(t) - \beta y(t)\\
 \displaystyle\frac{dz(t)}{dt} = \gamma x(t)^2 + \beta  y(t) - \rho z(t)
 \end{array}
 \label{eq:samplecompartmentalmodel}
\end{equation}
which can be put in the matrix form of (\ref{eq:statespace}). Due to the state-dependency of the rate flow between $x$ and $z$, the model is nonlinear. It is also an open system, since the sum of rate changes is non-zero, i.e., there is net flow in and out of the whole system (due to $u$ and $\rho$).
\begin{figure}[tb]
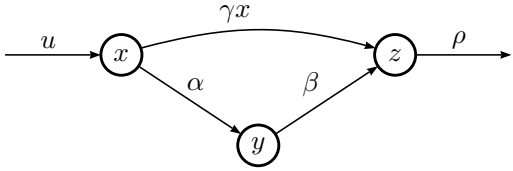

\MediumPicture\VCDraw{%
\begin{VCPicture}{(-1,-3)(3,2)}
\HideState
\State[0]{(0,0)}{X} \State[1]{(12,0)}{Y}
\ShowState
\State[x]{(3,0)}{A} \State[y]{(6,-2)}{B} \State[z]{(9,0)}{C}
\EdgeL{X}{A}{u}
\EdgeL{A}{B}{\alpha}
\EdgeL{B}{C}{\beta}
\ArcL{A}{C}{\gamma x}
\EdgeL{C}{Y}{\rho}
\end{VCPicture}
}
\caption{A sample compartmental model corresponding to the set of dynamic equations in (\ref{eq:samplecompartmentalmodel})}
\label{fig:compartmentalexample}
\end{figure}
\end{example}

\subsection{Mathematical epidemiology}
In order to model the propagation of epidemic diseases in a population, certain disease- and population-specific assumptions are required. The most common assumptions in this context include:
\begin{itemize}[leftmargin=*]
\item The diseases are contagious and transfer via contact.
\item A disease may or may not be fatal.
\item There may be births during the period of study, and the birth may (or may not) be congenitally transferred from the mother to the baby.
\item The disease can have an exposure period, during which the contaminants carry and spread the disease, but do not have visible symptoms.
\item Catching the disease may or may not result in short-term or long-term immunity. Depending on the case, the recovered patients can again become susceptible to the disease.
\item Interventions such as medication, vaccination, lockdown, quarantine and social distancing can change the pattern of propagation.
\end{itemize}
Let us consider an example, which is the basic model that we later extend for the COVIC-19 virus propagation pattern.

\subsubsection{The susceptible-infected-recovered model}
\label{ex:sir}
\begin{figure}[bt]
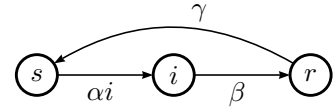

\MediumPicture\VCDraw{%
\begin{VCPicture}{(-4,-2)(6,2)}
\State[s]{(0,0)}{A} \State[i]{(3,0)}{B} \State[r]{(6,0)}{C}
\EdgeR{B}{C}{\beta}
\EdgeR{A}{B}{\alpha i}
\LArcR{C}{A}{\gamma}
\end{VCPicture} 
}
\caption{The basic susceptible-infected-recovered (SIR) model}
\label{fig:sir}
\end{figure}

A basic model used for modeling epidemic diseases without lifetime immunity is known as the \textit{susceptible-infected-recovered (SIR)} model \cite{anderson1979population,may1979population,brauer2012mathematical}. In this model, the total population of $N$ individuals exposed to an epidemic disease at each time instant $t$ is divided into three groups (each represented by a compartment): the susceptible group fraction denoted by $s(t)$, the infected group fraction denoted by $i(t)$, and the recovered group fraction denoted by $r(t)$ (the compartment variables are in fact the fraction of each group's population divided by $N$). Accordingly, the system is closed and we have
\begin{equation}
 s(t)+i(t)+r(t) = 1
 \label{eq:sirpopulation}
\end{equation}
A compartmental model for the propagation of the disease is shown in Fig.~\ref{fig:sir}. 
The compartmental representation of Fig.~\ref{fig:sir} is equivalent to the following set of differential equations:
\begin{equation}
\begin{aligned}
 &\frac{ds(t)}{dt}=-\alpha s(t)i(t) +\gamma r(t)\\
 &\frac{di(t)}{dt}= \alpha s(t) i(t) - \beta i(t)\\
 &\frac{dr(t)}{dt}= \beta i(t) -\gamma r(t)
 \end{aligned}
\label{eq:SIRequation}
\end{equation}
Accordingly, moving from the susceptible group to the infected group takes place at a rate that is proportional to the population of the infected and susceptible groups, with parameter $\alpha$. At the same time, infected individuals are assumed to recover at a constant rate of $\beta$. Finally, considering that the disease is not assumed to result in lifetime immunity of the subjects, the recovered individuals again return to the susceptible group at a fixed rate of $\gamma$. From (\ref{eq:SIRequation}), it is evident that 
\begin{equation}
\frac{ds(t)}{dt}+\frac{di(t)}{dt}+\frac{dr(t)}{dt}=0
\end{equation}
which is in accordance with (\ref{eq:sirpopulation}) and the fact that the system is assumed to be closed (no births or deaths have been considered).

\begin{figure}[bt]
\begin{subfigure}[Basic SIR with immunity]{\centering
\includegraphics[width=.47\textwidth]{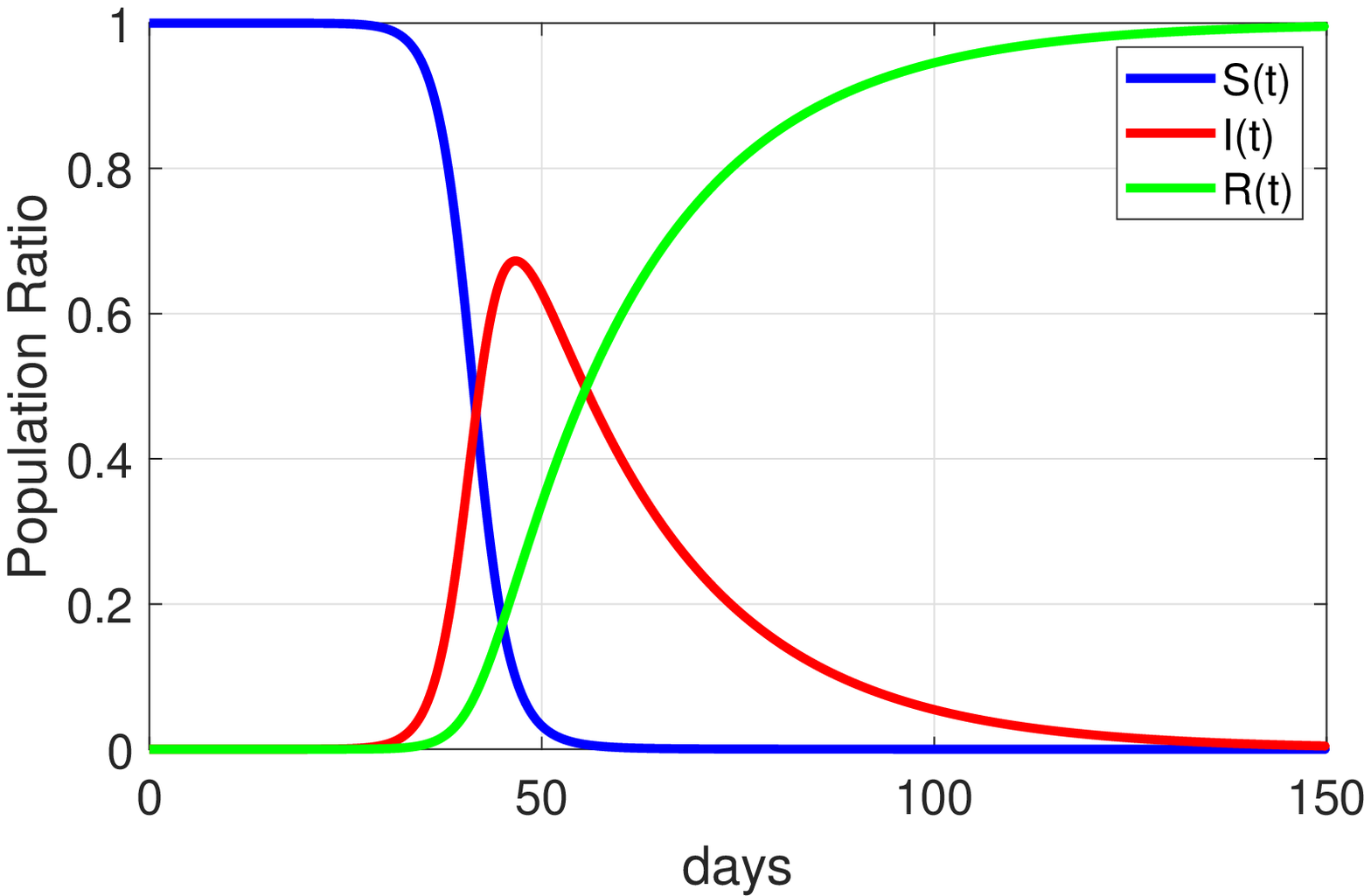}\label{fig:SIRSolutionimmune}}
\end{subfigure}
\begin{subfigure}[Basic SIR without immunity]{\centering\includegraphics[width=.47\textwidth]{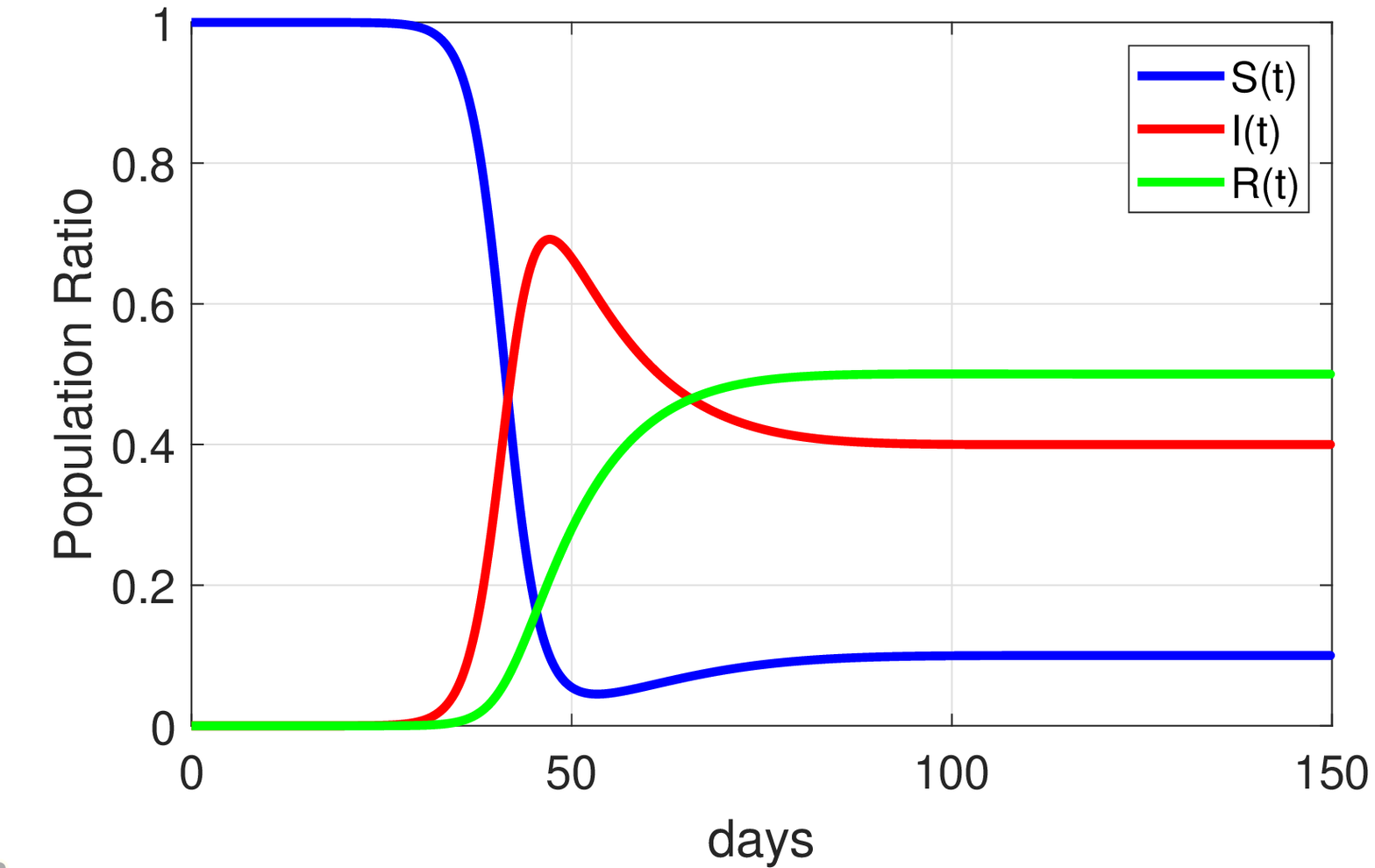}\label{fig:SIRSolutionnonimmune}}
\end{subfigure}
\caption{Simulation of a basic non-fatal SIR model with $\alpha$=0.5 and $\beta$=0.05 in two cases: a) $\gamma$=0.0 (lifetime immunity) and b) $\gamma$=0.04}
\label{fig:SIRSolution}	
\end{figure}

Assuming initial conditions for each group, the set of nonlinear equations (\ref{eq:SIRequation}) can be (numerically) solved to find the evolution of the population of each compartment over time. The numerical solution of a basic (non-fatal) SIR model is shown in  Fig.~\ref{fig:SIRSolution}, with and without lifetime immunity. The time-step for numerical discretizing of the differential equations of this simulation has been chosen to be $\Delta$=0.1 of a day. Notice how the outbreak of a disease that does not cause lifetime immunity (such as a typical flu), can result in a constant rate of illness throughout time, after its transient period. For widespread epidemic diseases, the healthcare strategists are interested in the slopes of $s(t)$, $i(t)$ and $r(t)$, rather than the total number of infected individuals (as it is currently the case for the COVID-19 coronavirus). The prolongation of the disease spread provides the better management of healthcare resources such as hospitalization, medication, healthcare personnel, etc.

For later reference, it is interesting to study the \textit{fixed-point} of the SIR model (where $\dot{s}(t)=\dot{i}(t)=\dot{r}(t)=0$). Equating the left sides of (\ref{eq:SIRequation}) with zero, it can be algebraically shown that if $\alpha, \gamma \neq 0$ (the non-immunizing case), the SIR model has only two fixed-points:
\begin{equation}
    \begin{array}{l}
         (s^*(t), i^*(t), r^*(t)) = (1, 0, 0)\\
         (s^*(t), i^*(t), r^*(t)) = \displaystyle(\frac{\beta}{\alpha}, I_0,\frac{\beta}{\gamma}I_0)
    \end{array}
\label{eq:sirfixedpoints}    
\end{equation}
where $I_0 \displaystyle\stackrel{\Delta}{=} \frac{\gamma(\alpha - \beta)}{\alpha(\gamma + \beta)}$. The first fixed-point corresponds to the lack of any infected cases, and the second corresponds to a persistent disease in the population, as illustrated in Fig.~\ref{fig:SIRSolutionnonimmune}. This situation is only reachable if $\beta < \alpha$, i.e., when the infection rate is greater than the recovery rate.

We can also verify whether or not the fixed-points are stable. Various methods can be used for this purpose. Perhaps, the most tangible approach is based on \textit{perturbation theory}. Simply stated, one can add small perturbations to the fixed-points of the system and check whether or not the perturbations are compensated by the system's dynamics by pushing the state vector back to its fixed-point. Accordingly, the first fixed-point in (\ref{eq:sirfixedpoints}) can be perturbed to:
\begin{equation}
(s(t), i(t), r(t)) = (1 - \epsilon, \epsilon, 0)
\label{eq:sirfixedpointsperturbed}    
\end{equation}
where $0 < \epsilon \ll 1$ is a small perturbation (e.g., equivalent to a single case of disease outbreak in a large population). Now replacing the perturbed point in (\ref{eq:SIRequation}) and neglecting second and higher order terms containing $\epsilon$, we obtain:
\begin{equation}
\begin{aligned}
 &\frac{ds(t)}{dt}=-\alpha (1 - \epsilon)\epsilon \approx -\alpha \epsilon < 0\\
 &\frac{di(t)}{dt}= \alpha (1 - \epsilon) \epsilon - \beta \epsilon \approx (\alpha - \beta)\epsilon\\
 &\frac{dr(t)}{dt}= \beta \epsilon > 0
 \end{aligned}
\label{eq:SIRequationperturbed}
\end{equation}
As a result, the first fixed-point is unstable, since due to the sign of the derivatives of the perturbed system, the system's dynamics drives the state vector away from the fixed-point (since the population of the susceptible group has a negative derivative). However, depending on whether $\alpha > \beta$ or not, the outbreak may or may not result in an increase in the infected population. Simply put, if the infection rate is greater than the recovery rate ($\alpha > \beta$) the disease would lead into an outspread; but if the recovery rate is faster than the infection rate ($\alpha < \beta$) the percentage of the infected population will remain close to zero. In either case, for a non-fatal non-immunizing disease, all individuals that become infected recover after a while and move to the recovered group and again go back to the susceptible group at a rate of $\gamma$. Note that a SIR model with a non-zero infected population fraction in steady-state, indicates that there is a constant flow between the compartments, i.e., people are constantly contaminated, recovered and again become susceptible to the disease.  

Perturbing the second fixed-point results in
\begin{equation}
\begin{aligned}
 &\frac{ds(t)}{dt}=-\alpha (\frac{\beta}{\alpha} - \epsilon)(I_0 + \epsilon) + \beta I_0 \approx \epsilon (\alpha I_0 - \beta)\\
 &\frac{di(t)}{dt}= \alpha (\frac{\beta}{\alpha} - \epsilon)(I_0 + \epsilon) - \beta I_0 \approx -\epsilon (\alpha I_0 - \beta)\\
 &\frac{dr(t)}{dt}= \beta (I_0 + \epsilon) -\beta(I_0 + \epsilon) = 0
 \end{aligned}
\label{eq:SIRequationperturbed2}
\end{equation}
In this case, depending on whether $(\alpha I_0 - \beta) > 0$ or not, the fixed-point may be stable or unstable.

For later use, we can show that during the outbreak of the SIR model ($s(t)\approx 1$), the number of infected cases follows an exponential pattern:
\begin{equation}
    i(t) \approx i(0)\exp[(\alpha - \beta)t]
\label{eq:SIRoutbreak}    
\end{equation}

\subsubsection{The fatal SIR model}
\begin{figure}[tb]
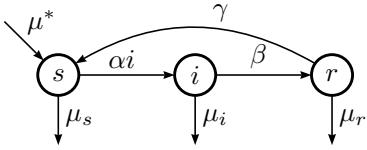

\MediumPicture\VCDraw{%
\begin{VCPicture}{(-4,-2)(6,2)}
\HideState
\State[0]{(-1.5,1.5)}{A0} \State[1]{(0,-2)}{A1} \State[2]{(3,-2)}{B1} \State[3]{(6,-2)}{C1}
\ShowState
\State[s]{(0,0)}{A} \State[i]{(3,0)}{B} \State[r]{(6,0)}{C}
\EdgeL{B}{C}{\beta}
\EdgeL{A}{B}{\alpha i}
\LArcR{C}{A}{\gamma}
\EdgeL{A0}{A}{\mu^*}
\EdgeL{A}{A1}{\mu_s}
\EdgeL{B}{B1}{\mu_i}
\EdgeL{C}{C1}{\mu_r}
\end{VCPicture} 
}
\caption{The susceptible-infected-recovered (SIR) model with birth and death rates}
\label{fig:endemicsir}
\end{figure}

A fatal version of the SIR model with rates of birth $\mu^*$ and with different death rates from the susceptible ($\mu_s$), infected ($\mu_i$) and recovered ($\mu_r$) groups is shown in Fig.~\ref{fig:endemicsir}. This system is no longer closed and its state equations can be written as follows:
\begin{equation}
\begin{aligned}
 \frac{ds(t)}{dt}&=\gamma r(t)-\alpha s(t)i(t) - \mu_s s(t) + \mu^*\\
 \frac{di(t)}{dt}&= \alpha s(t) i(t) - \beta i(t) - \mu_i i(t)\\
 \frac{dr(t)}{dt} &= \beta i(t)-\gamma r(t) -\mu_r r(t)
 \end{aligned}
\label{eq:SIRendemic}
\end{equation}

\subsection{The basic reproduction number (R0)\label{sec:R0}}
As noted before, the outbreak threshold of epidemiology models is known as the \textit{basic reproduction number}
$\mathcal{R}_0$. It is defined as the average number of secondary infections due to an infected individual hosted by a completely susceptible population \cite{hethcote2000mathematics,dietz1975transmission,kamgang2008computation}, \cite[Ch. 7]{diekmann2012mathematical}. The $\mathcal{R}_0$ during epidemic outbreak is generally greater than the average infections ($\mathcal{R}$) at any other time other than the outbreak. 

From the mathematical modeling perspective, a formal definition of $\mathcal{R}_0$ was first presented in \cite{van2002reproduction}. Consider the general dynamic representation of a compartmental model:
\begin{equation}
 \dot{\bm{x}}(t) = \mathbf{f}(\bm{x}(t)) 
\label{eq:statespacesimplified}    
\end{equation}
where $\bm{x}(t) = [x_1(t), x_2(t), \ldots, x_p(t), x_{p+1}(t), \ldots, x_n(t)]^T$ is the state vector (compartment variables), such that $x_1(t), \ldots, x_p(t)$ correspond to the infected compartments (exposed, infected, etc.), and $x_{p+1}(t), \ldots, x_n(t)$ are all the other variables (susceptibles, recovered, passed-away, etc.). We next partition each row of $\mathbf{f}(\cdot)$ as follows:
\begin{equation}
\begin{array}{l}
 \mathbf{f}(\bm{x}(t)) = \mathcal{F}(\bm{x}(t)) + \mathcal{V}(\bm{x}(t))
\end{array}
\label{eq:statespacepartitioned}    
\end{equation}
where $\mathcal{F}(\cdot)$ groups all the terms of $\mathbf{f}(\bm{x}(t))$, which correspond to \textit{new infections} (the portion of the population, which are either susceptible or had fully recovered, but are becoming exposed or infected due to contact with the exposed or infected). On the other hand, $\mathcal{V}(\cdot)$ groups all the other terms of the equations, including removals from the infected groups and other compartmental transitions.

The Jacobian of $\mathcal{F}(\cdot)$ and $\mathcal{V}(\cdot)$ are next calculated at the no infection fixed-point $\bm{x}^* = [0, 0, \ldots, 0, x_{p+1}^*, \ldots, x_n^*]^T$:
\begin{equation}
    \begin{array}{cc}
    \bm{\nabla}_{x}\mathcal{F}(x^*) = \begin{bmatrix}
    \mathbf{F} & \mathbf{0} \\
    \mathbf{0} & \mathbf{0}
    \end{bmatrix},
         &
\bm{\nabla}_{x}\mathcal{V}(x^*) = \begin{bmatrix}
    \mathbf{V} & \mathbf{0} \\
    \mathbf{J}_3 & \mathbf{J}_4
    \end{bmatrix}       
    \end{array}
\end{equation}
Finally, the reproduction number is defined as the spectral radius (leading eigenvalue) of the negative of the so-called \textit{next generation matrix} (NGM) $\mathbf{F}\mathbf{V}^{-1}$:
\begin{equation}
    \mathcal{R}_0 = \rho(-\mathbf{F}\mathbf{V}^{-1})
\end{equation}
which is proved to have the biological properties of the reproduction number for epidemic studies. 

In fact, while the threshold between stability and instability of an epidemic can be defined in various forms, only the definition based on $\mathcal{R}_0$ is biologically popular \cite{kamgang2008computation}. In \cite{van2002reproduction}, it is also shown how different partitionings of the state-space model can lead to different spectral radii; however, only the choice described in (\ref{eq:statespacepartitioned}) leads to the biologically meaningful definition of $\mathcal{R}_0$.

\begin{example}
\label{ex:SIRR0}
In the SIR model (\ref{eq:SIRequation}), if we replace $r(t) = 1 - s(t) - i(t)$ from (\ref{eq:sirpopulation}), the model reduces to:
\begin{equation}
\begin{aligned}
 &\frac{di(t)}{dt}= \alpha s(t) i(t) - \beta i(t)\\
&\frac{ds(t)}{dt}=-\alpha s(t)i(t) +\gamma [1 - s(t) - i(t)] 
 \end{aligned}
\label{eq:SIRequationreduced}
\end{equation}
Therefore,
\begin{equation}
    \begin{array}{ll}
    \mathcal{F} =
    \begin{bmatrix}
    \alpha s(t)i(t)\\
    0    
    \end{bmatrix},
         &
\mathcal{V} =
\begin{bmatrix}
    - \beta i(t)\\
    -\alpha s(t)i(t) +\gamma [1 - s(t) - i(t)]  
    \end{bmatrix}       
    \end{array}
\end{equation}
and at the fixed-point $\bm{x} = (0, 1)$
\begin{equation}
    \begin{array}{l}
    \bm{\nabla}_{x}\mathcal{F}(x^*) = \begin{bmatrix}
    \alpha s(t) & \alpha i(t) \\
    0 & 0
    \end{bmatrix}_{i(t) = 0, s(t) = 1}
         \\
\bm{\nabla}_{x}\mathcal{V}(x^*) = \begin{bmatrix}
    -\beta & 0 \\
    -\alpha s(t) - \gamma & -\alpha i(t) -\gamma
    \end{bmatrix}_{i(t) = 0, s(t) = 1}
    \end{array}
\end{equation}
which results in the reproduction number:
\begin{equation}
    \mathcal{R}_0 = \rho(-\mathbf{F}\mathbf{V}^{-1}) = \frac{\alpha}{\beta}
\label{eq:sirR0}    
\end{equation}
where we can see that the epidemic stability condition $\mathcal{R}_0 < 1$ is identical to the stability condition $\alpha < \beta$, found for the SIR model in Section \ref{ex:sir}.

\end{example}

Comparing (\ref{eq:sirR0}) and (\ref{eq:SIRoutbreak}) we notice that although the epidemic stability condition found from $\mathcal{R}_0$ is related to the outbreak exponent (slope of infection during outbreak), but they are not the same quantities.

In fact, a major drawback of the conventional definition of $\mathcal{R}_0$ using the NGM is that the discretization time (or generation period) is discarded in its definition and therefore, there is no direct analogy between the discrete-time and continuous-time outbreak behavior of the epidemic. Motivated by this fact and based on the analogy between the discrete-time and continuous-time models presented in Section \ref{sec:stochasticode}, we hereby propose an alternative definition of the reproduction number:
\mypropositionbox{An alternative definition of the reproduction number is
\begin{equation}
\tilde{\mathcal{R}}_0 = e^{\lambda_1 \Delta}
\label{eq:R0ProposedDefinition}
\end{equation}
where $\lambda_1$ is the real-part of the dominant eigenvalue of the dynamic model's Jacobian evaluated at the fixed-point of interest, and $\Delta$ is the generation time unit (or discretization period). Accordingly, for an irreducible dynamic model, $\tilde{\mathcal{R}}_0 < 1$ (or $\lambda_1 < 0$) and $\tilde{\mathcal{R}}_0 > 1$ (or $\lambda_1 > 0$) correspond to stable and unstable epidemic conditions, respectively.}

It is clear that for small generation time units $\Delta$ (small as compared with the compartmental model ``rate of variations'' in time), we have:
\begin{equation}
\tilde{\mathcal{R}}_0 \approx 1 + \lambda_1 \Delta
\end{equation}
The major advantage of the above definition for the reproduction number is that the time unit between generations appears in the definition. Therefore, the $\tilde{\mathcal{R}}_0$ of different epidemics that have been experimentally obtained from real-world data acquired with different generation time units become comparable with one another. Moreover, our studies on various epidemic models shows that the stability condition $\lambda_1 < 0$ (or $\tilde{\mathcal{R}}_0 < 1$) is exactly equivalent to the $\mathcal{R}_0 < 1$ condition obtained from the common definition of the basic reproduction number using the NGM. 
The mathematical proof of equivalence of the two conditions remains as future work.


\section{Proposed Epidemic Model I\label{sec:themodel}}
\begin{figure}[tb]
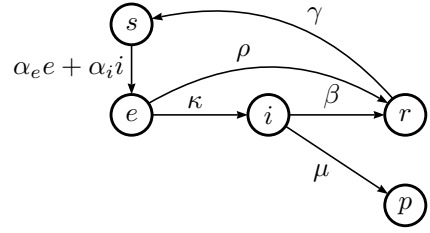

\MediumPicture\VCDraw{%
\begin{VCPicture}{(-2,-3)(6,2)}
\HideState
\State[3]{(6,-2)}{C1}
\ShowState
\State[s]{(3,2)}{A}
\State[e]{(3,0)}{B}
\State[i]{(6,0)}{C}
\State[r]{(9,0)}{D}
\State[p]{(9,-2)}{D1}
\EdgeL{C}{D}{\beta}
\EdgeL{B}{C}{\kappa}
\LArcL{B}{D}{\rho}
\EdgeR{A}{B}{\alpha_e e + \alpha_i i}
\LArcR{D}{A}{\gamma}
\EdgeR{C}{D1}{\mu}
\end{VCPicture} 
}
\caption{Proposed Model I: the fatal susceptible-exposed-infected-recovered (SEIR) model for coronavirus modeling with a unique recovery group}
\label{fig:endemicSEIR}
\end{figure}

Many infectious diseases are characterized by an incubation period between \textit{exposure} and the outbreak of \textit{clinical symptoms}. Subjects exposed to the infection are much more dangerous for the public as compared to the subjects showing clinical symptoms. The condition becomes more and more dangerous, with the increase of the isncubation rate. A well-known case is the HIV virus in its \textit{clinical latency} stage. The experience of the COVID-19 shows that a two-week incubation period can spread a virus worldwide and almost at any level of the society. \textit{Remember that any two of us are only six handshakes apart!\footnote{Cf. \url{https://en.wikipedia.org/wiki/Six_degrees_of_separation}}} For this reason, an additional compartment is added between the susceptibility and infection stages of the SIR model, which accounts for the asymptomatic exposed subjects. Moreover, since we are also interested in minimizing the mortality rate of the disease, a termination compartment is dedicated to the passed-away population. The variables of the model are therefore:
\begin{enumerate}[leftmargin=*]
    \item $s(t)$: The susceptible population fraction (the number of individuals in danger of being infected, divided by the total population).
    \item $e(t)$: The exposed population fraction (the number of individuals exposed to the virus but without having symptoms, divided by the total population).
    \item $i(t)$: The infected population fraction (the number of infected individuals with symptoms, divided by the total population).
    \item $r(t)$: The recovered population fraction (the number of recovered individuals, divided by the total population).    
    \item $p(t)$: The number of individuals that pass away due to the disease, divided by the total population).    
\end{enumerate}
Keeping in mind that
\begin{equation}
s(t) + e(t) + i(t) + r(t) + p(t) = 1
\label{eq:sumofstates}
\end{equation}
the proposed model and its compartmental representation are shown in equations (\ref{eq:SEIRPequation}) and Fig.~\ref{fig:endemicSEIR}.
\begin{equation}
\boxed{\begin{array}{l}
\text{\textbf{Model I:}}\\ \displaystyle\frac{ds(t)}{dt}=-\alpha_e s(t)e(t) -\alpha_i s(t)i(t) +\gamma r(t)\\
\displaystyle\frac{de(t)}{dt}= \alpha_e s(t)e(t) +\alpha_i s(t)i(t) - \kappa e(t) - \rho e(t)\\
\displaystyle\frac{di(t)}{dt}= \kappa e(t) - \beta i(t) - \mu i(t)\\
\displaystyle\frac{dr(t)}{dt}= \beta i(t) + \rho e(t) -\gamma r(t)\\
\displaystyle\frac{dp(t)}{dt}= \mu i(t)
 \end{array}}
\label{eq:SEIRPequation}
\end{equation}

In (\ref{eq:SEIRPequation}), similar to the classical SIR model, the interpretation of the nonlinear terms including $s(t)e(t)$ and $s(t)i(t)$ is that the rate of exposure to the virus is proportional the population of both the susceptible and exposed/infected subjects.

Note that the system closure constraint (\ref{eq:sumofstates}) gives an excess degree of freedom, which can be used to reduce the model order by replacing $s(t) = 1 - e(t) - i(t) -r(t) -p(t)$. This simplifies the compartmental model as follows:
\begin{equation}
\begin{aligned}
 \frac{de(t)}{dt}&= [1 - e(t) - i(t) -r(t) -p(t)][\alpha_e e(t) +\alpha_i i(t)] \\& - \kappa e(t) - \rho e(t)\\
 \frac{di(t)}{dt}&= \kappa e(t) - \beta i(t) - \mu i(t)\\
 \frac{dr(t)}{dt}&= \beta i(t) + \rho e(t) -\gamma r(t)\\
 \frac{dp(t)}{dt}&= \mu i(t)
 \end{aligned}
\label{eq:SEIRPequationReduced}
\end{equation}

\subsection{Measurements model}
Among the state variables of the proposed model, all except $e(t)$ are directly measurable (with potential errors). The measurements can be formulated in matrix form as follows:  
\begin{equation}
\left[\begin{aligned}
 I(t)\\
 R(t)\\
 P(t)
 \end{aligned}\right]=
\left[\begin{array}{cccc}
 0 & 1 & 0 & 0\\
 0 & 0 & 1 & 0\\
 0 & 0 & 0 & 1 
 \end{array}\right]
\left[\begin{aligned}e(t)\\i(t)\\r(t)\\p(t)
 \end{aligned}\right] +
\left[\begin{aligned}v_i(t)\\v_r(t)\\v_p(t)
 \end{aligned}\right]  
\label{eq:ObservationEquations}
\end{equation}
where $I(t)$ is the fraction of reported infections, $R(t)$ is the fraction of reported recoveries (both symptomatic and asymptomatic), $P(t)$ is the fraction of reported death tolls, and $\bm{v}(t) = [v_i(t), v_r(t), v_p(t)]^T$ is \textit{measurement noise}. The evident sources of measurement noises include: unavailable information regrading the exact population, intentional and unintentional misreported values, mis-classified reasons of death (especially for the elderly or subjects suffering from multiple health issues), and the marginal cases that may be unknown or misclassified for the healthcare system. Equation (\ref{eq:ObservationEquations}) can be written in more compact form as follows:
\begin{equation}
    \bm{y}(t) = \mathbf{C}\bm{x}(t) + \bm{v}(t)
\label{eq:observationequationmatrixform}    
\end{equation}
where $\bm{x}(t) = [e(t), i(t), r(t), p(t)]^T$ is the reduced state-vector. Although the variable $e(t)$ is not directly measurable from the available public data, we will show in Section \ref{sec:stateestimation} that under certain conditions, $e(t)$ can be estimated from the measurements. 

Note that in the above measurement model, it is assumed that $R(t)$ is the total recovery fraction of both the symptomatic and asymptomatic cases, assuming that the asymptomatic recoveries are measurable by (random or systematic) public tests over the population, such as the \textit{antibody} tests that have been conducted by some nations during the COVID-19 outbreak. In Section \ref{sec:therevisedmodel}, the model is modified to a more practical case, in which only the recoveries due to the symptomatic cases are measured.

\subsection{Model assumptions and level of abstraction}
The simplifying assumptions behind the proposed model are:
\begin{enumerate}[leftmargin=*]
    \item The model variables are assumed to be continuous in both amplitude and time. 
    \item Birth and natural deaths have been neglected. Therefore, other parameters leading to changes in the population are not considered. Neglecting the birth rate is also supported by the current findings that babies are not susceptible to this virus and to the best of our knowledge, no congenital transmissions of the virus from mothers to fetuses have been reported.
    \item In the current study, we do not distinguish between male and female subjects; although the current global toll of the virus suggests that men have been more vulnerable to the virus than women.
    \item Age ranges have not been considered; although we known that higher aged subjects are more vulnerable to the virus and countries have different \textit{population pyramids}.
    \item Moreover, in this primary version, we have not yet considered the possibility of vaccination.
    \item Geopolitical factors such as distance, country borders and continental differences have also been ignored. But considering that different countries have adopted customized countermeasures against the virus spread, the model parameters are fitted over country-level data.
\end{enumerate}

\subsection{Model parameters}
Having formed the model, we now explain its parameters and their relationship with real-world factors and clinical protocols. The techniques for estimating and fitting these parameters on real data is later detailed in Section \ref{sec:parameteridentification}.

Note that all the model parameters have the dimension of inverse time, to balance the left and right side dimensions of (\ref{eq:SEIRPequation}), and that the studied model is essentially based on an exponential law assumption, as detailed in Section \ref{sec:stochasticode}. Therefore, we can find rules of thumb for selecting the model parameters based on clinical facts and protocols. 

\begin{itemize}[leftmargin=*]
\item $\kappa$: The rate at which symptoms appear in exposed cases, resulting in transition from the exposed to the infected population. The selection of this parameter is according to the exponential law  detailed in Section \ref{sec:stochasticode}. Assume that we are dealing with an extremely contagious disease for which the healthcare decision makers have agreed on the above noted 99.75\% percentage as the target infection drop-out threshold, and advised 14 days of quarantine for the whole population. In that case, we can select $\kappa=6/14=0.43$ (inverse days) in our model. Apparently, there is a lot of simplifications in this discussion; the age range, the subject-specific body immune system features, the severeness of the virus and many more factors have been neglected. But it gives an idea about how the parameters can be tuned in practice, up to a reasonable order of magnitude. With this background, we now explain the interpretation of each parameter of the model.

We should add that wide screening policies adopted by certain countries are external factors that can significantly accelerate the identification of the infected cases. In this case, screening is a factor that increases $\kappa$.
\item $\alpha_i$: The contagion factor between the infected and susceptible populations, which is related to the contagiousness of the virus and social factors such as personal hygiene, population density and level of human interactions. In order to find the range (or order of magnitude) of this parameter, we can start with the contagion factors of more known viruses, such as flu and influenza, which are more or less influenced by the same spreading factors.
\item $\alpha_e$: The contagion factor between the exposed and susceptible populations. This parameter is logically far greater than $\alpha_i$, since in ordinary conditions (before lockdowns and quarantine), people rarely avoid contact with an asymptomatic individual; nor does the individual itself avoid interaction with others.
\item $\gamma$: The reinfection rate, or the rate of returning from the recovered group to the susceptible group. This happens for the cases that the body does not develop lifetime immunity after recovery, or the virus itself starts to mutate over time. This parameter is the inverse of the immunity rate of the virus. It is currently too early to comment about the immunity characteristics of the COVID-19 coronavirus\footnote{Refer to:\\ \url{https://www.who.int/docs/default-source/coronaviruse/who-china-joint-mission-on-covid-19-final-report.pdf}}. Although at least one case of reinfection soon after recovery has been reported, preliminary research have suggested short-term immunity of up to four months. 
\item $\beta$: The recovery rate of the infected cases. By considering the fourth equation in (\ref{eq:SEIRPequation}), we can denote the change in the number of hospitalized recoveries (or under control in any form, e.g., under home-care) by $r_h$, resulting in $r_h(t+\Delta)-r_h(t)\approx \Delta \beta i(t)$, where $\Delta$ is the time unit of approximation (for example 1 day). Therefore, the parameter $\beta$ can be approximated by dividing the daily recovery count of the population under study, by the total infected cases in the same day. In the real world, apart from the body strength of the infected subject in resisting against the virus, this parameter depends on the healthcare infrastructure of a country (hospitalization facilities, availability of medication, number of intensive care units, etc.).
\item $\rho$: The recovery rate of the exposed cases (the cases that are exposed, but recover without any symptoms). This parameter is not directly measurable from pure observations and requires lab-based experiments. However, we logically expect this parameter to be of the same order or greater than the parameter $\beta$ (the recovery rate of the infected population with symptoms). 
\item $\mu$: The mortality rate of the infected cases. By approximating the last equation in (\ref{eq:SEIRPequation}) by $p(t+\Delta)-p(t)\approx \Delta \mu i(t)$, where $\Delta$ is the time unit of approximation (for example 1 day), the parameter $\mu$ can be approximated by dividing the daily death toll by the total infected cases in the same day. As with $\beta$, the mortality of the virus itself, the immune system of the subjects, and the medical infrastructure are important factors that influence the parameter.
\item $e_0$: The initial exposed population (seed).
\end{itemize}

By studying the above factors, we can see that the only parameters of the model that can be changed in the short-term (before the development of long-term solutions such as vaccination, medication, improvement of hospitalization facilities, etc.), is to reduce the infection rates by minimizing human contacts (social distancing), or to apply public screening. These are the two policies, which have been adopted worldwide.

\subsection{Fixed-point analysis}
As with the basic SIR model presented in Example \ref{ex:sir}, the fixed-point(s) of the model can be sought by letting the left hand sides of all equations in (\ref{eq:SEIRPequation}) equal to zero. Accordingly, assuming that all the model parameters are nonzero, the only fixed-point is the no-disease case ($i(t)=e(t)=r(t)=0$):
\begin{equation}
(s^*(t), e^*(t), i^*(t), r^*(t), p^*(t)) = (1 - p_0, 0, 0, 0,p_0)
\label{eq:seirpfixedpoints}    
\end{equation}
where $0 \leq p_0 \leq 1$ is the steady-state total death fraction. The stability of this fixed-point can be addressed by perturbing the fixed-point with a minor exposure $\epsilon$ (which can correspond to a single new exposed case in the real world):
\begin{equation}
(s(t), e(t), i(t), r(t), p(t)) = (1 - p_0 - \epsilon, \epsilon, 0, 0, p_0)
\label{eq:seirpfixedpointsperturbed}    
\end{equation}
Putting this point in the state dynamics (\ref{eq:SEIRPequation}), we find:
\begin{equation}
\begin{aligned}
\frac{ds(t)}{dt}&=-\alpha_e (1 - p_0 - \epsilon)\epsilon\approx-\alpha_e(1 - p_0)\epsilon < 0\\
 \frac{de(t)}{dt}&= \alpha_e (1 - p_0 - \epsilon)\epsilon - \kappa \epsilon - \rho \epsilon\approx (\alpha_e - \alpha_e p_0 - \kappa - \rho)\epsilon\\
 \frac{di(t)}{dt}&= \kappa \epsilon > 0\\
 \frac{dr(t)}{dt}&= \rho \epsilon > 0\\
 \frac{dp(t)}{dt}&= 0
 \end{aligned}
\label{eq:SEIRTequationperturbed}
\end{equation}
which is unstable, i.e., the system's dynamics drives it away from the fixed-point in the direction of reducing the healthy cases, resulting in further infection. A more rigorous study of the system stability conditions is presented in the following sections using eigenanalysis.

\subsection{Model analysis during outbreak}
\label{sec:outbreakmodel}
Let us study the model during the initial outbreak of the epidemic, when the infection toll is still much smaller than the total population. For instance, suppose that a country has 100,000 of exposed or infected cases, which is indeed significant for any country, as it is far beyond the available number of intensive care unit beds of even the most developed countries\footnote{See for example:\\ \url{https://link.springer.com/article/10.1007/s00134-012-2627-8/tables/2}}. But for a 100 million population country, such an exposure/infection toll is only 0.1\% of the total population. Therefore, during the primary phases of the disease spread, the model can be simplified by assuming that the susceptible population is almost constant ($s(t) \approx 1$) and $ds(t)/dt \approx 0$, regardless of the other parameters of the model. This assumption practically implies that the total population is not important during an epidemic outbreak (in low percentages of infection), resulting in

\myresultbox{In low percentages of infection, the performance of epidemic control policies of states, countries, and regions should not be evaluated by normalizing the infection/recovery/death tolls to their total population; but rather the net values should be compared.}

This result has also been approved in previous research based on model fitting on data from several epidemic diseases, showing that the disease spread is considerably independent of the total population size \cite{hethcote2000mathematics}.

Under this assumption, (\ref{eq:SEIRPequation}) is simplified to the linear set of equations:
\begin{equation}
\left[\begin{aligned}
 \frac{de(t)}{dt}\\
 \frac{di(t)}{dt}\\
 \frac{dr(t)}{dt}\\
 \frac{dp(t)}{dt}
 \end{aligned}\right]\approx
\left[\begin{array}{cccc}
 \alpha_e - \kappa -\rho & \alpha_i & 0 & 0 \\
 \kappa & -\beta - \mu & 0 & 0\\
 \rho & \beta & -\gamma & 0\\
 0 & \mu & 0 & 0
 \end{array}\right]
\left[\begin{aligned}e(t)\\i(t)\\r(t)\\p(t)
 \end{aligned}\right] 
\label{eq:SEIRPequationSimplified}
\end{equation}
Defining $\bm{x}(t) = [e(t), i(t), r(t), p(t)]^T$, (\ref{eq:SEIRPequationSimplified}) can be written in matrix form:
\begin{equation}
    \frac{d}{dt}\bm{x}(t) = \mathbf{A}\bm{x}(t)
\label{eq:lineardynamicform}    
\end{equation}
where $\mathbf{A}$ is the 4$\times$4 state matrix on the right hand side of (\ref{eq:SEIRPequationSimplified}). Equation (\ref{eq:lineardynamicform}) can be solved for an arbitrary initial condition, such as $\bm{x}(0) = (e_0, 0, 0, 0)$. The characteristic function of this linear system is:
\begin{equation}
    |\lambda \mathbf{I} - \mathbf{A}| = \lambda (\lambda + \gamma) [\lambda^2 + (\beta+\mu-\delta)\lambda -\delta(\beta+\mu) -\kappa \alpha_i] = 0
\end{equation}
where $\delta\stackrel{\Delta}{=}(\alpha_e - \kappa - \rho)$. Therefore the system's eigenvalues are:
\begin{equation}
\begin{array}{l}
\lambda_1=\displaystyle\frac{\delta-\beta-\mu + \sqrt{(\delta+\beta+\mu)^2 + 4\kappa \alpha_i}}{2}\\
\lambda_2=\displaystyle\frac{\delta-\beta-\mu -\sqrt{(\delta+\beta+\mu)^2 + 4\kappa \alpha_i}}{2}\\    \lambda_3 = 0,\quad \lambda_4 = -\gamma
    \end{array}
\label{eq:eigenvalues}    
\end{equation}
which are all real-valued. Moreover, it is straightforward to check that $\lambda_1 > \delta > \lambda_2$. The eigenvectors corresponding to each eigenvalue are:
\begin{equation}
\begin{array}{l}
    \mathbf{v}_1=\displaystyle k_1[1, \frac{\lambda_1 - \delta}{\alpha_i}, \frac{\rho\alpha_i+\beta(\lambda_1-\delta)}{\alpha_i(\lambda_1+\gamma)}, \frac{\mu(\lambda_1-\delta)}{\alpha_i\lambda_1}]^T\\
    \mathbf{v}_2=\displaystyle k_2[1, \frac{\lambda_2 - \delta}{\alpha_i}, \frac{\rho\alpha_i+\beta(\lambda_2-\delta)}{\alpha_i(\lambda_2+\gamma)}, \frac{\mu(\lambda_2-\delta)}{\alpha_i\lambda_2}]^T\\
    \mathbf{v}_3 = [0,0,0,k_3]^T, \quad \mathbf{v}_4 = [0,0,k_4,0]^T    
    \end{array}
\label{eq:eigenvectors}  
\end{equation}
where $k_1$, $k_2$, $k_3$ and $k_4$ are arbitrary constants. The general form of the solution of the compartmental variables is a summation of exponential terms with the above exponential rates and eivenvectors:
\begin{equation}
    \bm{x}(t) = \displaystyle\sum_{k=1}^4 a_k e^{\lambda_k t} \mathbf{v}_k
\end{equation}
Specifically, after some algebraic simplifications, we can calculate the infected and exposed populations as follows:
\begin{equation}
\begin{array}{l}
i(t) = \displaystyle \frac{e_0(\lambda_1-\delta)(\delta-\lambda_2)}{\alpha_i(\lambda_1-\lambda_2)} [\exp{(\lambda_1 t)} - \exp{(\lambda_2 t)}]\\
e(t) = \displaystyle \frac{e_0}{\lambda_1-\lambda_2} [ (\lambda_1-\delta)\exp{(\lambda_2 t)} + (\delta-\lambda_2)\exp{(\lambda_1 t)}]
\end{array}
\label{eq:infectiousToll}    
\end{equation}
From the last equation in (\ref{eq:SEIRPequation}), it is clear that the death toll will not stop before $i(t) = 0$. Also from (\ref{eq:infectiousToll}), we can see that since $\lambda_1$ is the dominant eigenvalue, the steady-state behavior and whether or not $i(t)$ and $e(t)$ diverge from or converge to zero, depends on the sign of $\lambda_1$. The necessary and sufficient condition for the linearized system's stability (stopping the death doll) is $\lambda_1 < 0$, which simplifies to $\kappa\alpha_i + \delta(\beta+\mu) < 0$, or:
\begin{equation}
\boxed{
\kappa\alpha_i < (\kappa + \rho -\alpha_e)(\beta+\mu)}
\label{eq:stabilitycondition}    
\end{equation}
A sufficient condition that guarantees this property is when $\alpha_i=\alpha_e=0$. The condition $\alpha_i=0$ implies that the susceptible group avoids contact with the infected ones. However, the second condition ($\alpha_e=0$) is difficult to fulfill in the real-world, since the exposed group do not have any symptoms. This is why social distancing is required to enforce $\alpha_e\approx 0$ and to permit all the exposed subjects to move to the infected group without infecting new individuals, after which the asymptomatic group can be all considered clear of the disease. Another practical case is when $\alpha_i\approx 0$ (healthy people avoid contact with the infected) and $\kappa + \rho > \alpha_e$ (the rate of recovery of the exposed or the appearance of their symptoms is faster than the rate of new exposures). This condition is fulfilled by social distancing and lockdown (isolation of even the asymptomatic cases for a certain period).

However, if none of the above conditions are fulfilled and $\lambda_1 > 0$, the number of exposed and infected cases increases exponentially at a rate of $\lambda_1$. In this case, with fixed system parameters, the infection rate rises exponentially up to a point at which the linear approximation does no longer hold. This practically translates into:
\myresultbox{During an exponential outbreak of an epidemic ($\lambda_1 > 0$), the system is unstable and without enforcing temporary lockdowns, social distancing and quarantine of the infected cases (resulting in the model parameter changes), the exponential increase in the number of infected subjects continues to a point where a significant percentage of the population is infected.}

Using the method detailed in Section \ref{sec:R0}, we can further show that for the epidemic model (\ref{eq:SEIRPequation}), the reproduction number (spectral radius of the NGM) is equal to:
\begin{equation}
    \mathcal{R}_0 = \frac{\alpha_e(\beta + \mu) + \kappa\alpha_i}{(\kappa + \rho)(\beta+\mu)}
\label{eq:SEIRPR0}    
\end{equation}
Apparently, $\mathcal{R}_0 < 1$ exactly simplifies to the stability condition in (\ref{eq:stabilitycondition}), when $\lambda_1 < 0$. 

\myresultbox{Under countermeasures, the model eigenvalues change and $\lambda_1$ (the dominant eigenvalue of the linearized dynamic model) is the single  parameter that can be tracked as a score for evaluating how good countermeasures such as social distancing and quarantine are performing.}

Considering that the death toll $p(t)$ is composed of the same exponential terms as the infected cases in
(\ref{eq:infectiousToll}), the above result is indeed disturbing.

It is also interesting to observe from (\ref{eq:infectiousToll} ) that the population of the different compartments of the model is only linearly proportional to the initial exposed population size $e_0$\footnote{Note that the COVID-19 is believed to have started from a single case.}. Therefore, for a large population (at the level of a populated city or country), the initial infected seed size is not as important as the other model parameters that influence the exponential behavior of the model (such as the social contact rates). Therefore:
\myresultbox{The initial seed size is not the most critical parameter for epidemic management. Regions with smaller initial seeds of infected/exposed cases may end up with a higher infected and death toll depending on their infection rates, defined by factors such as human-contact rate and personal hygiene.}

Another interesting property is to check the ratio between the number of infected (which is measurable in the real world) and the number of exposed (which is not directly measurable). From (\ref{eq:infectiousToll}), we can find\footnote{The numerator and denominator of (\ref{eq:IERatio}) have been multiplied by $\exp(-\delta t)$ to obtain the simplified form.}:
\begin{equation}
    \frac{i(t)}{e(t)} = \frac{\exp(\tilde{\lambda}_1 t) - \exp(-\tilde{\lambda}_2 t)}{\alpha_i[\tilde{\lambda}_1^{-1}\exp(\tilde{\lambda}_1 t) + \tilde{\lambda}_2^{-1}\exp(-\tilde{\lambda}_2 t)]}
    \label{eq:IERatio}
\end{equation}
where $\tilde{\lambda}_1\stackrel{\Delta}{=}\lambda_1 - \delta$ and $\tilde{\lambda}_2\stackrel{\Delta}{=}\delta - \lambda_2$ are both positive. Therefore, when the terms containing $\exp(-\tilde{\lambda}_2 t)$, which is a decaying exponential, vanish and the epidemic model is still in its linear phase ($i(t) \ll s(t)$ or $s(t)\approx 1$), the ratio can be approximated by:
\begin{equation}
    \frac{i(t)}{e(t)} \rightarrow \frac{\tilde{\lambda}_1}{\alpha_i} \quad \text{for $t \gg \tilde{\lambda}_2^{-1}$ and $i(t) \ll s(t)$}
    \label{eq:IERatioSimplified}
\end{equation}
which gives the following practical result:
\myresultbox{During the primary phases of an epidemic outbreak (when the number of contaminated cases has an exponential growth, but the percentage of the infected individuals to the total population is still small), the number of exposed subjects can be approximated by $e(t)\approx \alpha_i\tilde{\lambda}_1^{-1}i(t)$, permitting its estimation from $i(t)$.}

\subsection{Repeated waves of epidemic}
The peaks of the infected group population, and its potential repetition in time, is important from the strategic viewpoint \cite{feng2007final}. These points correspond to local or global extremums of $i(t)$, which mathematically correspond to where $di(t)/dt=0$ in (\ref{eq:SEIRPequation}), i.e., where $i(t)\displaystyle=\kappa e(t)/(\beta+\mu)$. It can be shown that this leads to a reduced order set of nonlinear dynamic equations, which can be solved for the remaining variables $[s(t), e(t), r(t), p(t)]^T$. The simulations demonstrated in the sequel, show that the infected population can have multiple local peaks over time, with recurrent behaviors, proving that:
\myresultbox{The epidemic disease can repeat pseudo-periodically over time (in later seasons or years) and turn into a persistent disease in the long term. The amplitude and time gap of the infection peaks depends on the model parameters.}

This behavior has been observed in previous pandemics, such as the 1918 pandemic influenza, known as the \textit{Spanish flu}, where three pandemic waves of infection have been observed within an interval of a few months\footnote{Cf. \url{https://en.wikipedia.org/wiki/Spanish_flu}}. A mathematical study of sustained oscillations of similar compartmental models has been studied in previous research \cite{may1979population,gonccalves2011oscillations}.

\section{Proposed Epidemic Model II\label{sec:therevisedmodel}}
\begin{figure}[tb]
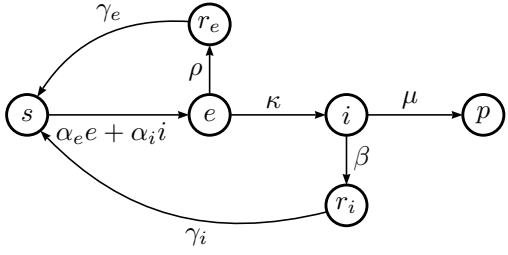

\MediumPicture\VCDraw{%
\begin{VCPicture}{(-2,-3)(6,2)}
\ShowState
\State[s]{(0,0)}{A}
\State[e]{(4,0)}{B}
\State[i]{(7,0)}{C}
\State[p]{(10,0)}{D}
\State[r_e]{(4,2)}{E}
\State[r_i]{(7,-2)}{F}
\EdgeL{C}{F}{\beta}
\EdgeL{B}{C}{\kappa}
\EdgeL{B}{E}{\rho}
\EdgeR{A}{B}{\alpha_e e + \alpha_i i}
\LArcR{E}{A}{\gamma_e}
\LArcL{F}{A}{\gamma_i}
\EdgeL{C}{D}{\mu}
\end{VCPicture} 
}
\caption{Proposed Model II: an extension of the fatal SEIR model for coronavirus modeling, with separate recovery groups from exposure and infection}
\label{fig:endemicSEIRiRe}
\end{figure}

The proposed model can be extended from various aspects. One such extension is to separate the recoveries from exposure from the recoveries from infection. The advantage of this separation is that in practice, the subjects that recover without any symptoms may only be identified by broad public screening, which is not very practical for a large population. While the infected individuals that recover are already known for the healthcare system and are easy to monitor. Based on this idea, the proposed extension of the model is as shown in Fig.~\ref{fig:endemicSEIRiRe}. Accordingly, the variables $i(t)$, $p(t)$ and $r_i(t)$ (the recoveries from infection) are the variables that can be observed and reported by the healthcare units. The dynamic system corresponding to this model is:
\begin{equation}
\boxed{\begin{array}{l}
\text{\textbf{Model II:}}\\ \displaystyle\frac{ds(t)}{dt}=-\alpha_e s(t)e(t) -\alpha_i s(t)i(t) +\gamma_e r_e(t)+\gamma_i r_i(t)\\
\displaystyle\frac{de(t)}{dt}= \alpha_e s(t)e(t) +\alpha_i s(t)i(t) - \kappa e(t) - \rho e(t)\\
\displaystyle\frac{di(t)}{dt}= \kappa e(t) - \beta i(t) - \mu i(t)\\
\displaystyle\frac{dr_e(t)}{dt}= \rho e(t) -\gamma_e r_e(t)\\
\displaystyle\frac{dr_i(t)}{dt}= \beta i(t) -\gamma_i r_i(t)\\
\displaystyle\frac{dp(t)}{dt}= \mu i(t)
 \end{array}}
\label{eq:SEIRPequationModel2}
\end{equation}
subject to $s(t) + e(t) + i(t) + r_e(t) + r_i(t) + p(t) = 1$, which can again be used to reduce the model order by omitting one of the model variables (e.g., $s(t)$). In the latter case, we assume that the observed variables are $I(t)$, $R_i(t)$ and $P(t)$, resulting in the following observation model:
\begin{equation}
\left[\begin{aligned}
 I(t)\\
 R_i(t)\\
 P(t)
 \end{aligned}\right]=
\left[\begin{array}{ccccc}
 0 & 1 & 0 & 0 & 0\\
 0 & 0 & 0 & 1 & 0\\
 0 & 0 & 0 & 0 & 1 
 \end{array}\right]
\left[\begin{aligned}e(t)\\i(t)\\r_e(t)\\r_i(t)\\p(t)
 \end{aligned}\right] +
\left[\begin{aligned}v_i(t)\\v_r(t)\\v_p(t)
 \end{aligned}\right]  
\label{eq:ObservationEquationsModel2}
\end{equation}
which can be written in compact form, as in (\ref{eq:observationequationmatrixform}).

Similar to the first model, under the assumption of low fraction of infection (during epidemic outbreak) and omitting the variable $s(t)$, (\ref{eq:SEIRPequationModel2}) simplifies to:
\begin{equation}
\resizebox{.97\hsize}{!}{$\left[\begin{aligned}
 \frac{de(t)}{dt}\\
 \frac{di(t)}{dt}\\
 \frac{dr_e(t)}{dt}\\
 \frac{dr_i(t)}{dt}\\
 \frac{dp(t)}{dt}
 \end{aligned}\right]\approx
\left[\begin{array}{ccccc}
 \alpha_e - \kappa -\rho & \alpha_i & 0 & 0 & 0 \\
 \kappa & -\beta - \mu & 0 & 0 & 0\\
 \rho & 0 & -\gamma_e & 0 & 0\\
 0 & \beta & 0 & -\gamma_i & 0\\
 0 & \mu & 0 & 0 & 0
 \end{array}\right]
\left[\begin{aligned}e(t)\\i(t)\\r_e(t)\\r_i(t)\\p(t)
 \end{aligned}\right]$}
\label{eq:SEIRPequationSimplifiedModel2}
\end{equation}
Defining the state vector $\bm{x}(t) = [e(t), i(t), r_e(t), r_i(t), p(t)]^T$, (\ref{eq:SEIRPequationSimplifiedModel2}) can be written in a matrix form similar to (\ref{eq:lineardynamicform}), where $\mathbf{A}$ is now the 5$\times$5 state matrix on the right hand side of (\ref{eq:SEIRPequationSimplifiedModel2}) and solved for an arbitrary initial condition. The characteristic function of this linear system is:
\begin{equation}
    |\lambda \mathbf{I} - \mathbf{A}| = \lambda (\lambda + \gamma_e)(\lambda + \gamma_i) [\lambda^2 + (\beta+\mu-\delta)\lambda -\delta(\beta+\mu) -\kappa \alpha_i]
\end{equation}
where as in the first model $\delta\stackrel{\Delta}{=}(\alpha_e - \kappa - \rho)$, resulting in the following eigenvalues:
\begin{equation}
\begin{array}{l} \lambda_1=\displaystyle\frac{\delta-\beta-\mu + \sqrt{(\delta+\beta+\mu)^2 + 4\kappa \alpha_i}}{2}\\
\lambda_2=\displaystyle\frac{\delta-\beta-\mu -\sqrt{(\delta+\beta+\mu)^2 + 4\kappa \alpha_i}}{2}\\
    \lambda_3 = 0,\quad \lambda_4 = -\gamma_e,\quad \lambda_5 = -\gamma_i\\
    \end{array}
\label{eq:eigenvaluesModel2}    
\end{equation}
In this case, the eigenvectors corresponding to each eigenvalue are:
\begin{equation}
\begin{array}{l}
    \mathbf{v}_1=\displaystyle k_1[1, \frac{\lambda_1 - \delta}{\alpha_i},
    \frac{\rho}{\lambda_1+\gamma_e},
    \frac{\beta(\lambda_1-\delta)}{\alpha_i(\lambda_1+\gamma_i)},
    \frac{\mu(\lambda_1-\delta)}{\alpha_i\lambda_1}]^T\\
    \mathbf{v}_2=\displaystyle k_2[1, \frac{\lambda_2 - \delta}{\alpha_i},
    \frac{\rho}{\lambda_2+\gamma_e},
    \frac{\beta(\lambda_2-\delta)}{\alpha_i(\lambda_2+\gamma_i)},
    \frac{\mu(\lambda_2-\delta)}{\alpha_i\lambda_2}]^T\\
    \mathbf{v}_3 = [0,0,0,0,k_3]^T, \quad \mathbf{v}_4 = [0,0,k_4,0,0]^T\\
    \mathbf{v}_5 = [0,0,0,k_5,0]^T
    \end{array}
\label{eq:eigenvectorsModel2}  
\end{equation}
where $k_1$, $k_2$, $k_3$, $k_4$ and $k_5$ are arbitrary constants. The first three eigenvalues are identical to Model I. Therefore, the outbreak properties such as exposure/infection rates, stability conditions and reproduction number are exactly the same as the first model, as derived in (\ref{eq:infectiousToll}), (\ref{eq:stabilitycondition}), and (\ref{eq:SEIRPR0}).

\section{Epidemic trend estimation, model observability and control\label{sec:stateestimation}}
The ability to estimate the future trend of the epidemic pattern is extremely important from the strategic perspective. This requires the accurate estimation of the compartmental model variables from reported statistics of the virus spread. The trend estimation requirements are addressed in the sequel for both the linearized and general form of both of the proposed models. As noted before, these results can be used to design an estimation scheme, based on for example a Kalman filter, for estimation and prediction of the current and future trends, in presence of inaccurate infection tolls. Note that although the variable $e(t)$ is not directly measurable from the available public data, one may seek whether of not this variable can be indirectly estimated from the other measurements (assuming that the other model parameters are known).

\subsection{Observability during outbreak (low fraction of infection)}
During the epidemic outbreak (in the low fraction of infection case), the linearized versions of Model I and Model II, namely (\ref{eq:SEIRPequationSimplified}) and (\ref{eq:SEIRPequationSimplifiedModel2}) hold, respectively. Using the notion of \textit{observability} from system theory \cite{Kailath1980,terrell2009stability}, for the model matrix pair $(\mathbf{A}, \mathbf{C})$ the observability matrix is defined as follows:
\begin{equation}
\bm{\mathcal{O}}_k =
    \begin{bmatrix}
    {\mathbf{C}}_{k}\\
    {\mathbf{C}}_{k}{\mathbf{A}}_{k}\\
    \vdots\\
    {\mathbf{C}}_{k}{\mathbf{A}}_{k}^{n - 1}
    \end{bmatrix}
\label{eq:observability}    
\end{equation}
If $\bm{\mathcal{O}}_k$ has rank $n$ (the number of state variables, in either of the proposed models), all the state variables of the linearized models (\ref{eq:SEIRPequationSimplified}) and (\ref{eq:SEIRPequationSimplifiedModel2}) are observable at the outputs, which means that they can be estimated from the observations in finite time. It is straightforward to numerically calculate (\ref{eq:observability}) for arbitrary choices of the model parameters of Model I and Model II, and to check that none of its columns are linearly dependent\footnote{Matlab has a function called \texttt{obsv} for calculating the observability matrix, from the matrix pair $(\mathbf{A}, \mathbf{C})$.}. Therefore, all the state variables, including $e(t)$, are observable and may be estimated using state estimation techniques such as the Kalman filter (as the optimal linear estimator).

\subsection{Observability in the general case}
In the general case, where the number of exposed/infected cases exceeds several percents of the population, the variations in $s(t)$ is no longer negligible and one should refer to the original nonlinear compartmental models  (\ref{eq:SEIRPequation}) and (\ref{eq:SEIRPequationModel2}). In this case, the observability rank test (\ref{eq:observability}) can be checked for the matrix pair $(\bm{\mathcal{A}}(t), \mathbf{C})$, where the matrix $\mathbf{C}$ is similar to the linearized case in (\ref{eq:ObservationEquations}) and $\bm{\mathcal{A}}(t)$ is the Jacobian of the nonlinear model (\ref{eq:SEIRPequationReduced}), with respect to the entries of the reduced state-vector $\bm{x}(t) = [e(t), i(t), r(t), p(t)]^T$. The entries of the Jacobian have been listed in Appendix \ref{sec:jacobians}; some of which are time-dependent. Due to the non-evident form of the observability matrix of this case, the proof of observability of the matrix pair $(\bm{\mathcal{A}}(t), \mathbf{C})$ in its general case is cumbersome. However, it is simple to check this property numerically for arbitrary values of the model parameters. We have tested this property for our later shown simulated results, as part of the source codes provided online at \cite{EpidemicModelingCodes}. Therefore, we can state the following result in both the general and linear approximated case:
\myresultbox{Although the number of exposed cases of the population is not directly measurable, if the model parameters are known (or accurately estimated), the number of exposed cases can also be estimated from the other observations.}

\subsection{Epidemic control and model controllability\label{sec:controllability}}
The proposed Models I and II, do not have external inputs. Therefore, without changing the model parameters, there is no control mechanism for the epidemic and one may only verify the conditions under which the system is \textit{internally stable}, i.e., the effect of epidemic outbreaks would vanish, remain bounded, or result in an exponential outbreak. However, interventions such as social distancing, quarantine, medication, vaccination, etc., can be considered as control inputs that change the model parameters. 
%

\section{Model parameter identification in fixed and socially confined scenarios\label{sec:parameteridentification}}
An important step in making the proposed model useful in practice is to fit its parameters on real data. As noted before, the general compartmental model in (\ref{eq:statespace}) depends on the parameter vector $\bm{\theta}(t)$, which is generally time dependent (relies on social contact, hygiene, etc.). In order to fit the model parameters various methods are available in the literature. We study two approaches, which can be applied to our problem of interest.
\subsection{Constrained least squares parameter estimation}
A general formulation for parameter identification is to use constrained weighted least squares (CWLS) estimation. This approach becomes equivalent to the maximum likelihood estimate, if the measurement noises belong to specific families of probability distributions (such as the Gaussian distribution). Nevertheless, the CWLS is more generic as it only attempts in finding the parameter vector that minimizes a quadratic error cost function between the model and measurements (without any probabilistic priors on the origin of the model or measurement noises). We can specifically refer to the CWLS formulation proposed in \cite{gabor2015robust} and other prior research, which have been specifically developed for nonlinear dynamic models. Accordingly, for the general dynamic model (\ref{eq:statespace}), if we define the modeling error function
\begin{equation}
    \bm{e}(t) \stackrel{\Delta}{=} \bm{y}(t) - \mathbf{g}(\bm{x}(t); \bm{\theta}(t), t), 
\label{eq:errorfunction}    
\end{equation}
assuming temporarily that the model parameters $\bm{\theta}(t) = \bm{\theta}$ are fixed, the problem of parameter identification can be formulated as follows:
\begin{equation}
    \begin{array}{l}
    \hat{\bm{\theta}} = \displaystyle\argmin_{\bm{\theta}} {\Tr{\mathbb{E}[\bm{e}(t) \mathbf{W} \bm{e}(t)^T]}}\\
    \text{subject to:}\\
    \bm{\theta}_{\text{min}} \leq \bm{\theta} \leq \bm{\theta}_{\text{max}},\\
    \displaystyle\dot{\bm{x}}(t) = \mathbf{f}(\bm{x}(t), \bm{w}(t); \bm{\theta}, t), \quad \bm{x}(0) = \bm{x}_0\\
    \end{array}
\label{eq:CWLSproblem}    
\end{equation}
Where $\mathbb{E}(\cdot)$ denotes averaging over time, $\Tr(\cdot)$ denotes matrix trace, $\bm{\theta}_{\text{min}}$ and $\bm{\theta}_{\text{max}}$ define the lower and upper bounds of the model parameters dictated by physical constraints, and $\mathbf{W}$ is the inverse of the covariance matrix of the measurement noise vector $\bm{v}(t)$ (if available; otherwise can be set to identity for an unweighted version). This problem is in the form of nonlinear CWLS for which a variety of stable numerical solvers exist. Refer to \cite{gabor2015robust} for a survey of methods and \cite[Ch. 5]{moeller2003mathematical} for methods specific to dynamic systems. In (\ref{eq:CWLSproblem}), since the temporal averaging is performed over all time samples, the procedure is only applicable to offline model fitting. Now if $\bm{\theta}(t)$ is time variant, an adaptive version of the above algorithm can be used, by averaging the error cost function over $t_0 \leq \tau \leq t$, resulting in a sample-wise updates of the parameter vector. In the following subsection, we propose an alternative approach for the time variant case, which is more flexible, does not require nonlinear least-squares solvers, and simultaneously estimates the model state variables.

\subsection{An extended Kalman filter for joint parameter and variable estimation}
A well-known method for adaptive estimation of dynamic system parameters is to consider the (possibly) time-varying parameters of the model as additional state variables with presumed dynamics and to estimate them at the same time or in parallel with the original state vector. For this, let us assume that all the parameters of our base epidemic model (\ref{eq:SEIRPequation}) are time-variant, resulting in the parameter vector $\bm{\theta}(t)=[\alpha_i(t),\alpha_e(t), \kappa(t), \beta(t), \rho(t), \mu(t), \gamma(t)]^T$, that should be estimated from real-world epidemic data. From the modeling perspective, this vector can have some deterministic or stochastic fluctuations. For example, let us assume that the model parameters are a simple Wiener process (Brownian motion) plus deterministic inputs to model social interventions:
\begin{equation}
\begin{array}{ll}
\displaystyle\frac{d \alpha_i(t)}{dt} =  u_{\alpha_i}(t) + w_i(t)&
\displaystyle\frac{d \alpha_e(t)}{dt} = u_{\alpha_e}(t) + w_e(t)\\
\displaystyle\frac{d \kappa(t)}{dt} = u_{\kappa}(t) + w_{\kappa}(t) &
\displaystyle\frac{d \beta(t)}{dt} = u_{\beta}(t) + w_{\beta}(t)\\
\displaystyle\frac{d \rho(t)}{dt} = u_{\rho}(t) + w_{\rho}(t)&
\displaystyle\frac{d \mu(t)}{dt} = u_{\mu}(t) + w_{\mu}(t)\\
\displaystyle\frac{d \gamma(t)}{dt} = u_{\gamma}(t) + w_{\gamma}(t)
\end{array}
\label{eq:paramdynamics}
\end{equation}
where $\bm{w}(t) = [w_i(t), w_e(t), w_{\kappa}(t), w_{\beta}(t), w_{\rho}(t), w_{\mu}(t), w_{\gamma}(t)]^T$ is zero-mean white noise acting as \textit{process noise}, and $u_{\alpha_i}(t)$ and $u_{\alpha_e}(t)$ are the deterministic inputs due to social intervention that change the level of social contact (considered for modeling the effect of social distancing and quarantine). For example, the effect of lockdown applied at time $t = t_0$ can be modeled by combinations of functions of the form:
\begin{equation}
    \begin{array}{l}
         u_{\alpha_i}(t) = a_i - b_i e^{-c_i (t - t_0)}u(t - t_0)  \\
         u_{\alpha_e}(t) = a_e - b_e e^{-c_e (t - t_0)}u(t - t_0) 
    \end{array}
\end{equation}
where $c_i$ and $c_e$ define the speed of lockdown application, and $a_i$, $b_i$, $a_e$ and $b_e$ define the range of lockdown effect. The effect of lockdown termination can be modeled in a similar manner. In any case, as detailed in Appendix \ref{sec:jacobians}, the above dynamics can be \textit{state augmented} with the compartmental model dynamics in (\ref{eq:SEIRPequation}) and (\ref{eq:ObservationEquations}) to form an augmented model, which can be tracked using an extended Kalman filter (EKF). The implementation of an EKF requires the Jacobian matrix of the state augmented model with respect to all the elements of the augmented state vector $\tilde{\bm{x}}(t) \stackrel{\Delta}{=}[\bm{x}(t); \bm{\theta}(t)]$ (where $;$ denotes column-wise stacking), as detailed in the appendix. The details of the EKF is beyond the scope of the current study, which is focused on the modeling aspects of the problem. Nevertheless, the algorithmic steps of the discretized version of the compartmental model are presented in Appendix \ref{sec:EKF}. An implementation of the EKF for epidemic model parameter and state vector tracking will be provided online in the Git repository of the project \cite{EpidemicModelingCodes}. The interested reader is referred to classical textbooks for the required signal processing and algorithmic details of the EKF and its extensions \cite[Ch. 5 and 6]{Hay01}, \cite{GrewalAndrews01}, \cite[P. 191]{Gel74}, \cite[Appendix 9.A and 9.B]{jazwinski1970stochastic}.

Note that a great advantage of using the EKF for estimating the model states and parameters is that in addition to estimating the values, it also provides \textit{confidence intervals} for the estimates (which is an advantage of all Bayesian inference/estimation methods). In other words, under the given assumptions on the process and measurement noise statistics, one can for example determine how accurate the number of infected and exposed cases have been estimated. In the context of interest, such confidence intervals permit healthcare strategists to have an idea about the accuracy of the estimated values and appropriate timing for lockdown and quarantine management. 



\section{Estimation pandemic trend and reproduction rate from noisy reports}
\label{sec:trendestimation}
In Section \ref{sec:outbreakmodel}, it was shown that during the outbreak of a pandemic, the number of new cases has an exponential growth, corresponding to the dominant eigenvalue of the dynamic system. Tracking this exponential rate over time from daily reported cases under social distancing and intervention plans is of practical importance. A major limitation in this context is the errors, inconsistencies and delays of the reported cases over weekdays, weekends and other holidays. This has been a recurrent issue during the COVID-19 pandemic. Therefore, most analysts prefer to monitor the three-day or seven-day moving average of the reported cases. We will show that an extended Kalman filter/smoother can be effectively used for the simultaneous prediction/filtering of the daily reports and also the estimation of the reproduction rate under NPI or other social distancing factors. For this, we consider the following model, for the number of new cases for an arbitrary region (during outbreak):
\begin{equation}
\begin{array}{l}
x(t+\Delta) = x(t)\exp[\Delta\lambda(t)]  + w(t) \\
\lambda(t + \Delta) = \displaystyle\sigma \tanh\{\frac{1}{\sigma}[\alpha \lambda(t) + v(t) + h(\mathbf{p}(t))]\}\\
y(t) = x(t) + n(t)
\end{array}
\label{eq:overallmodel}
\end{equation}
where:
\begin{itemize}
    \item $t$: is time in days.
    \item $\Delta$: reporting time period ($\Delta = 1$ for daily reports)
    \item $x(t)$: the \textit{true} number of new cases (to be estimated)
    \item $y(t)$: the reported number of new cases (erroneous)
    \item $\lambda(t)$: the pandemic spread exponent
    \item $w(t)$: exponential model deviation error (white noise)
    \item $v(t)$: autoregressive model driving noise (white noise)
    \item $\alpha$: autoregressive model rate ($0 < \alpha < 1$)
    \item $n(t)$: observation/reporting error due to weekends, holidays, counting errors, etc.
    \item $\mathbf{p}(t)\in\mathbb{R}^{L}$: a vector of $L$ quantitative NPI indexes such as the Oxford COVID-19 Government Response Tracker (OxCGRT) \cite{OxCGRT2020}, and other indexes such as holidays, seasonal temperature, etc.
    \item $h(\cdot)$: the NPI influence function, which is a mapping from the NPIs to $\lambda(t)$, to be found by learning mechanisms such as \textit{least absolute shrinkage and selection operator} (LASSO) or \textit{long short-term memory} (LSTM) recurrent neural networks. Note that $h(\mathbf{p}(t))$ acts as an \textit{exogenous input} for the dynamic model. Considering that the NPIs are planned by authorities and are rather consistent for a certain period (e.g. several weeks) until they become effective in reducing the number of new cases, $h(\mathbf{p}(t))$ commonly acts as a step-response that in the steady-state changes the DC (average) level of $\lambda(t)$. 
    \item $\sigma$: the exponential rate saturation value set to the maximum absolute value of the anticipated $\lambda(t)$ for a specific region or country. In fact, the $\tanh$ saturation function in (\ref{eq:overallmodel}), smoothly saturates $\lambda(t)$ and guarantees model stability. By selecting $\sigma$ several times greater than the standard deviation of $\lambda(t)$ over the past several months of each region, the saturation function will only saturate the outliers and will not influence normal variations of $\lambda(t)$.
\end{itemize}

Note that in (\ref{eq:overallmodel}), following the proposed definition in (\ref{eq:R0ProposedDefinition}), the exponential growth factor can be related to the time-varying reproduction rate as follows: $\tilde{\mathcal{R}}(t) = e^{\lambda(t) \Delta}$. 

Having developed the model, the trend estimation consists of two phases: the \textit{endogenous} part in absence of any NPI, and the \textit{exogenous} part due to the NPIs, which are detailed below.

\subsection{Endogenous trend prediction}
In absence of NPIs, we may assume $h(\mathbf{p}(t)) = 0$ (or constant) in (\ref{eq:overallmodel}). Defining the state vector $\bm{s}(t) = [x(t), \lambda(t)]^T$, the first two equations in (\ref{eq:overallmodel}) serve as smooth differentiable nonlinear state equations and the third equation is the linear observation equation. In our online repository, we have implemented an extended Kalman filter (EKF) and a fixed-interval extended Kalman smoother (EKS) based on this endogenous model. The algorithm is summarized in Appendix \ref{sec:EKFtrendtracking}. During the training phase, the EKS is used for smoothing the noisy daily reports and to estimate the exponential rate $\lambda(t)$. The same model is used in an EKF during the prediction phase of the challenge, but without having any new case reports. The EKF and EKS may even be used for estimating the missing dates (such as weekends and holidays that countries do not report exact cases). The entries of the covariance matrices of the EKF/EKS, which are integral elements of all Bayesian estimators including the EKF/EKS are also used to calculate confidence intervals for the estimated values.

\subsection{Endogenous prediction}
During the training period, $\hat{\lambda}(t)$ (an estimate of $\lambda(t)$) is obtained by the EKF for an arbitrary region/country. Next, from the second equation in (\ref{eq:overallmodel}), the residual terms corresponding to the exogenous NPI input and autoregressive noise is obtained. This term is given to regression models such as LASSO or LSTM as the target output, together with the regional NPIs and other indexes to learn $h(\cdot)$. The overall algorithm for predicting the new cases is summarized in Algorithm \ref{alg:proposedpredictor}.

\begin{algorithm}[tbh]
\caption{Proposed algorithm for new cases forecasting \label{alg:proposedpredictor}}
\begin{algorithmic}[1]
\STATE \textit{Training phase:} apply the EKS on reported new cases to estimate $\hat{\lambda}(t)$, the exponential rate of pandemic spread.
\STATE Use regression algorithms such as LASSO or LSTM to learn $h(\cdot)$, the regression map from the NPIs to $\hat{\lambda}(t)$.
\STATE \textit{Forecasting phase:} the trained model $h(\cdot)$ is used in the EKF to forecast future values of the new cases and its exponential rate, together with the confidence intervals obtained from the covariance matrices of the EKF.
\end{algorithmic}
\end{algorithm}

\subsection{Alternative methods for exponential rate estimation}
Beyond the Kalman filter based approach, we have also implemented three other methods for estimating the rate of growth during the offline training phase. Assuming that $\lambda(t)$ is rather constant over a small windows of $N$ days, it can be estimated by solving the following nonlinear least squares error problem:
\begin{equation}
\lambda(t) = \argmin_{\lambda} \sum_{t' = t - N + 1}^t\{y(t') - x(t) \exp[(t' - t)\lambda]\}^2
\label{eq:nlinfit}
\end{equation}
which can be numerically solved using optimization toolboxes such as \texttt{nlinfit} in Matlab or the \texttt{scipy.optimize} package for Python. An approximate workaround for the nonlinear problem in (\ref{eq:nlinfit}) is to apply a linear regression on the logarithm of the new cases time-series over a sliding window of $N$ days. Another method is an empirical geometric mean of the new cases ratios. A Matlab implementation of these methods is provided in our online repository.

\begin{figure}[tb]
\begin{subfigure}[The new case]{
\includegraphics[width=.95\columnwidth]{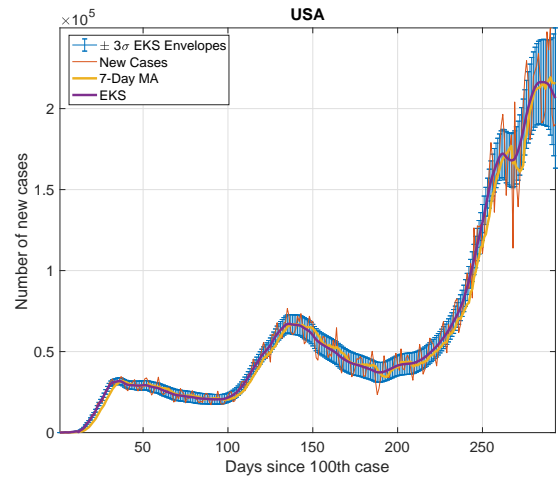}\label{fig:NewCasesEKSUSA}}
\end{subfigure}\\
\begin{subfigure}[The exponential growth]{\includegraphics[width=.95\columnwidth]{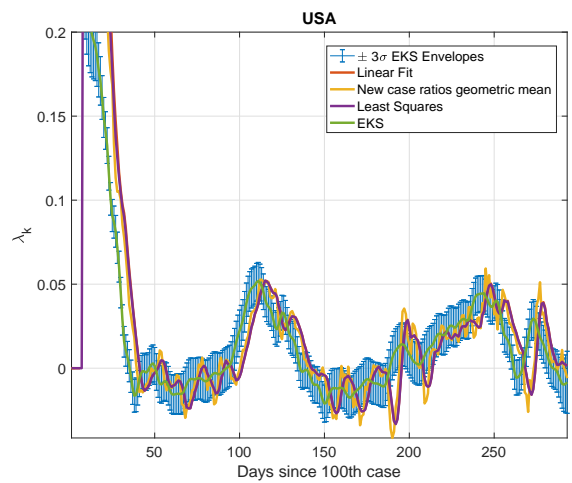}\label{fig:LambdaEKSUSA}}
\end{subfigure}\\
\caption{Tracking the trend of new cases and the exponential growth using the proposed extended Kalman smoother on daily reported cases of the US, since the 100th case report. The raw noisy daily reports have been adopted from the Oxford COVID-19 Government Response Tracker (OxCGRT) project \cite{OxCGRT2020}.}
\label{fig:EKSSampleResult}	
\end{figure}

A sample result of tracking the trend of new cases and the exponential growth using the hereby proposed extended Kalman smoother on real daily reported cases of the US since the 100th case report is shown in Fig.~\ref{fig:EKSSampleResult}. For simplicity, the effect of NPIs have not been considered in this example.

\section{Simulated results}
We have carried-out several simulations with different sets of parameters, resulting in the different scenarios explained in previous sections.

\begin{example}[Life-time immune case]
\label{example:ScanarioA}
The first scenario is illustrated in Fig.~\ref{fig:SEIRPSimulationsA}, where we have considered a single virus outbreak in a 84 million population, with constant parameters $\alpha_e = 0.65$, $\alpha_i = 0.005$, $\kappa = 0.05$, $\rho = 0.08$, $\beta = 0.1$, $\mu = 0.02$ and $\gamma = 0$, simulated over one year. The assumption $\gamma = 0$ implies that the disease has been assumed to develop lifetime immunity, therefore there is no return from the recovered to the susceptible compartment. In Fig.~\ref{fig:SEIRPSimulationsA}, we can observe the constant ratio between $i(t)$ and $e(t)$ during the exponential outbreak (when $i(t) \ll s(t)$ and before reaching the nonlinear phase of the model), which approves the relationship derived in (\ref{eq:IERatioSimplified}).
\begin{figure}[b]
\centering
\includegraphics[trim=1cm 0.0cm 1cm 0.5cm, clip=true, width=\columnwidth]{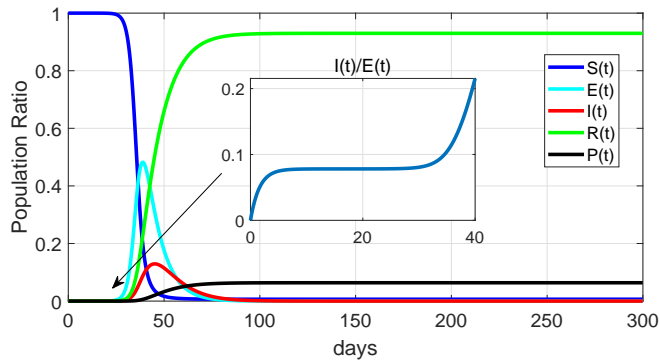}
\caption{Scenario A: Simulation of the fatal SEIR model with parameters $\alpha_e = 0.65$, $\alpha_i = 0.005$, $\kappa = 0.05$, $\rho = 0.08$, $\beta = 0.1$, $\mu = 0.02$ and $\gamma = 0$, over 10 months. Notice the constant $i(t)/e(t)$ ratio during the exponential outbreak, which is approximately equal to $\alpha_i\tilde{\lambda}_1^{-1}=0.078$.}
\label{fig:SEIRPSimulationsA}	
\end{figure}
\end{example}

\begin{example}[Recurrent epidemic]
\label{example:ScanarioB}
The second scenario is illustrated in Fig.~\ref{fig:SEIRPSimulationsB}. All the model parameters are identical to Example \ref{example:ScanarioA}, except for the re-susceptibility rate (loss of immunity) that is now $\gamma = 0.001$. It is interesting to see that in this scenario, the model has recurrent decaying peaks over time, similar to the aforementioned 1918 Spanish flu.
\begin{figure}[tb]
\begin{subfigure}[in five months]{
\includegraphics[width=.95\columnwidth]{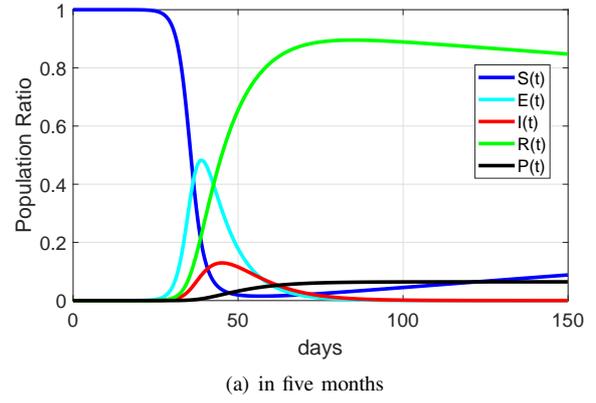}\label{fig:SEIRPSimulationsB1}}
\end{subfigure}\\
\begin{subfigure}[in ten years]{\includegraphics[width=.95\columnwidth]{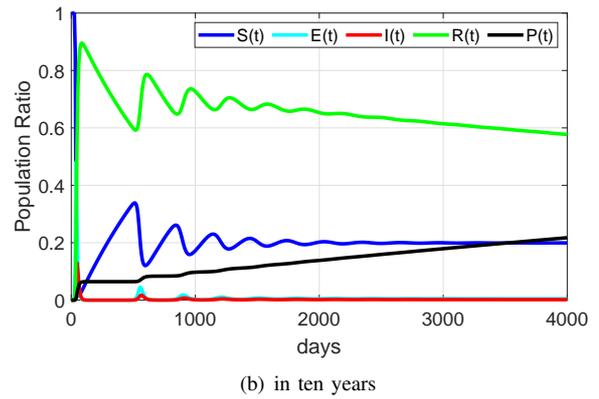}\label{fig:SEIRPSimulationsB2}}
\end{subfigure}\\
\begin{subfigure}[$I(t)$ and $E(t)$ in ten years]{\includegraphics[width=.95\columnwidth]{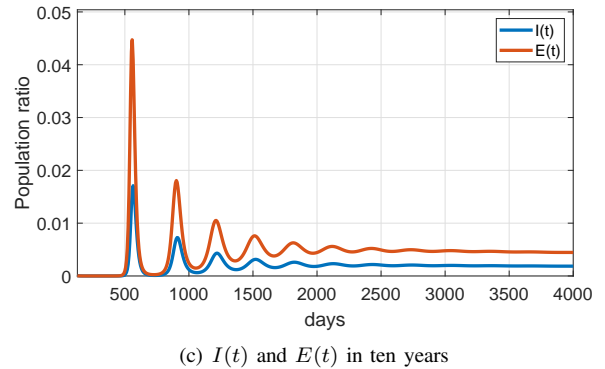}\label{fig:SEIRPSimulationsB3}}
\end{subfigure}
\caption{Scenario B: Simulation of the fatal SEIR model with parameters $\alpha_e$ = 0.65, $\alpha_i$ = 0.005, $\kappa = 0.05$, $\rho$ = 0.08, $\beta$ = 0.1, 
$\mu$ = 0.02, $\gamma$ = 0.001, in (a) five months and (b, c) ten years}
\label{fig:SEIRPSimulationsB}	
\end{figure}
\end{example}

\begin{example}[Short lockdown period]
\label{example:ScanarioC}
The next scenario is illustrated in Fig.~\ref{fig:shortquarantine}. This scenario corresponds to the case where a one month lockdown is applied by the government to identify the exposed and infected cases. The parameters of the model before quarantine are $\alpha_e = 0.6$, $\alpha_i = 0.005$, $\kappa = 0.05$, $\rho = 0.08$, $\beta = 0.1$, $\mu = 0.02$ and $\gamma = 0.001$. In the 30th day after the initial outbreak, the one month lockdown is applied, during which $\alpha_e$ and $\alpha_i$ are reduced to 0.1 and 0.001, respectively, while keeping the other parameters unchanged. After one month, $\alpha_i$ remains 0.001 (people keep distance with the infected ones), but $\alpha_e$ is increased to 0.4 (much more than the quarantine period, but two-third of the original social contact factor). We can see that with this scenario, which corresponds to an insufficient quarantine period, the population of the infected and exposed peaks have decreased; but the quarantine does not significantly change the mortality toll after five months. But why? Because, after the quarantine period, there is still a minor fraction of the population that is exposed, and this very small seed can re-initiate the virus spread. We therefore come to the following result:
\myresultbox{Imposing quarantines is effective in delaying and reducing the infection population peaks; but is insufficient in the long term. Social distancing and other measures should remain for a long period after the initial quarantine, to make the number of contaminated subjects equal to ``zero.''}

\begin{figure}[tbh]
\centering
\includegraphics[width=.9\columnwidth]{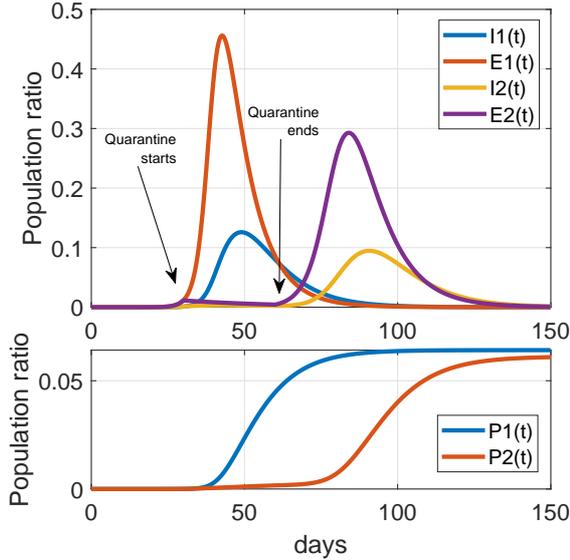}
\caption{Scenario C: Simulation of the fatal SEIR model with a quarantine period between day 30 and 60. $E1(t)$, $I1(t)$ and $P1(t)$ correspond to the case without quarantine and $E2(t)$, $I2(t)$ and $P2(t)$ correspond to the case with quarantine. Refer to the text for further details and the switching of the parameters.}
\label{fig:shortquarantine}	
\end{figure}
\end{example}

\begin{example}[Healthcare system saturation]
\label{example:ScanarioD}
Our next scenario corresponds to the case where the healthcare system reaches its maximum capacity or break-point, due to limited test kits, medication, hospitalization, excessive fatigue or mortality of the healthcare personnel, economic breakdowns, etc. This phenomenon is the worst feared case for pandemic strategists and it can be modeled at various levels. Suppose that we model it in its simplest form and assume that the recovery and mortality rates of the model change as functions of the number of infected cases:
\begin{equation}
    \begin{array}{l}
    \beta(t) = (\beta_s - \beta_0) h(i(t)) + \beta_0\\
    \mu(t) = (\mu_s - \mu_0) h(i(t)) + \mu_0    
    \end{array}
\end{equation}
where $\beta_0$ and $\mu_0$ are the recovery and mortality rate parameters before healthcare system saturation, $\beta_s$ ($\beta_s \ll \beta_0$) and $\mu_s$ ($\mu_s \gg \mu_0$) are the recovery and mortality rates after saturation, and $h(i(t))$ is a saturation function such that $h(0) \approx 0$ and $h(\infty) \approx 1$. For example, the hyperbolic tangent with a slope parameter $\sigma$ is a reasonable and common choice for modeling such phenomena:
\begin{equation}
    h(i) = \frac{1}{2}[1 + \tanh{(\frac{i - i_0}{\sigma})}]  
\end{equation}
where $i_0$ is the infection break-point of the healthcare system. The above choice is also beneficial for parameter optimization and tracking, due to the smoothness and differentiability of the hyperbolic tangent function.

A simulation of this scenario is illustrated in Fig.~\ref{fig:healthcarebreakpoint}. The parameters of the model are $\alpha_e = 0.6$, $\alpha_i = 0.005$, $\kappa = 0.05$, $\rho = 0.08$, and $\gamma = 0.001$. The healthcare system breakdown parameters are $\beta_0 = 0.1$, $\mu_0 = 0.02$, $\beta_s = 0.01$, $\mu_s = 0.2$, $\sigma = 1$, $i_0 = 0.1$ (although in reality, the healthcare system is expected to reach its breakpoint far below this infection rate). We can observe how the death rate significantly increases as the system reaches its break-point.
\begin{figure}[tbh]
\centering
\includegraphics[width=.9\columnwidth]{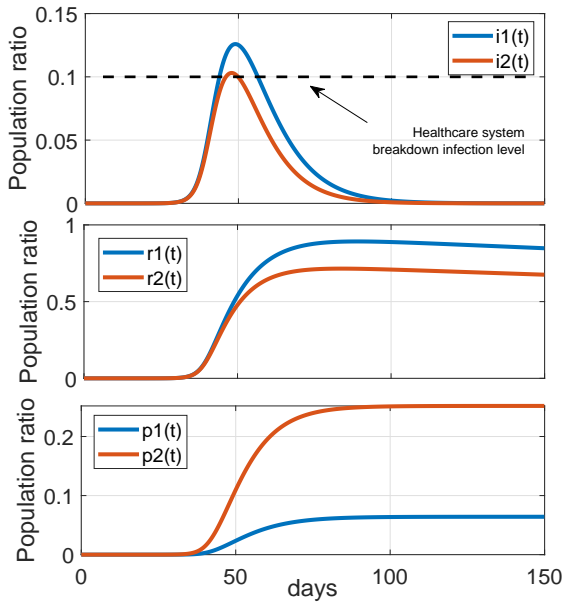}
\caption{Scenario D: Simulation of the fatal SEIR model when the healthcare system reaches its breakdown point at $i(t)= 0.1$. The curves $i1(t)$, $r1(t)$, and $p1(t)$ are the non-saturated cases and $i2(t)$, $r2(t)$, and $p2(t)$ correspond to the saturated cases.}
\label{fig:healthcarebreakpoint}	
\end{figure}
\end{example}

\section{Conclusion and Future Work}
In this study we considered some of the properties of epidemic diseases from a mathematical and signal processing perspective, by using a compartmental model for the propagation of the epidemic disease. It was shown that the model is not stable around its fixed-point (the no-infection case). 
We should indeed be proactively worried about the instability of our societies to such epidemic outbreaks; since a fatal virus with an initial seed as small as a single subject worldwide can trigger the avalanche of pandemic waves, killing many people worldwide, as the COVID-19 coronavirus has shown. Note however that from the modeling and dynamics perspective, there is nothing specific to the coronavirus, apart from its specific parameters. In fact, humans have always been living in such a metastable condition and researchers from all domains need to synergize to think a way through this condition.

The research will be continued and extended from various aspects in the future versions. Specifically, by fitting the model over real data, the prediction of infection and mortality rates under quarantine and vaccination, and the study of the recurrent pattern of the epidemic disease over time. Non-exponential law infection distributions have also been considered in the literature \cite{feng2007epidemiological}. This can be an interesting track of research for the study of epidemic diseases, including the family of coronaviruses. 

\appendix

\subsection{Compartmental Model Jacobians\label{sec:jacobians}}
Following (\ref{eq:SEIRPequationReduced}) and the Wiener process model for the model parameters presented in (\ref{eq:paramdynamics}), the most general form of the proposed compartmental model with time-varying parameters is as follows
\begin{equation}
\begin{array}{l}
 \displaystyle\frac{de(t)}{dt}= [1 - e(t) - i(t) -r(t) -p(t)][\alpha_e(t) e(t) +\alpha_i(t) i(t)] \\ \qquad\qquad- \kappa(t) e(t) - \rho(t) e(t)\\
 \displaystyle\frac{di(t)}{dt}= \kappa(t) e(t) - \beta(t) i(t) - \mu(t) i(t)\\
 \displaystyle\frac{dr(t)}{dt}= \beta(t) i(t) + \rho(t) e(t) -\gamma(t) r(t)\\
 \displaystyle\frac{dp(t)}{dt}= \mu(t) i(t)\\
\displaystyle\frac{d \alpha_i(t)}{dt} = \frac{d u_i}{dt}(t) + w_i(t)\qquad
\displaystyle\frac{d \alpha_e(t)}{dt} = \frac{d u_e}{dt}(t) + w_e(t)\\
\displaystyle\frac{d \kappa(t)}{dt} = w_{\kappa}(t)\qquad
\displaystyle\frac{d \beta(t)}{dt} = w_{\beta}(t)\qquad
\displaystyle\frac{d \rho(t)}{dt} = w_{\rho}(t)\\
\displaystyle\frac{d \mu(t)}{dt} = w_{\mu}(t)\qquad
\displaystyle\frac{d \gamma(t)}{dt} = w_{\gamma}(t) 
\end{array}
\label{eq:SEIRPequationReducedTVARParlams}
\end{equation}
If we define the variable and parameter augmented state vector:
\begin{equation}
\begin{array}{rl}
\bm{x}(t)\stackrel{\Delta}{=}&[e(t), i(t),r(t), p(t),\\&\alpha_i(t),\alpha_e(t),\kappa(t), \beta(t), \rho(t), \mu(t), \gamma(t)]^T
\end{array}
\end{equation}
which can be written in the compact state-space form (\ref{eq:statespace}). The Jacobian of this form is defined
\begin{equation}
\bm{\mathcal{A}}(t) \displaystyle\stackrel{\Delta}{=} \frac{\partial \mathbf{f}(\cdot)}{\partial \bm{x}}\in \mathbb{R}^{11\times 11}
\end{equation}
with the following non-zero entries (we have dropped the time index of all variables and parameters for better readability):
\begin{equation}
\begin{array}{l}
\bm{\mathcal{A}}_{1,1} = \alpha_e [1 - 2e - i - r - p] - \alpha_i i - \kappa - \rho\\
\bm{\mathcal{A}}_{1,2} = \alpha_i[1 - e - 2 i - r - p] -\alpha_e e\\
\bm{\mathcal{A}}_{1,3} = -[\alpha_e e +\alpha_i i]\quad
\bm{\mathcal{A}}_{1,4} = -[\alpha_e e +\alpha_i i]\\
\bm{\mathcal{A}}_{1,5} = i [1 - e - i - r - p]\quad
\bm{\mathcal{A}}_{1,6} = e [1 - e - i - r - p]\\
\bm{\mathcal{A}}_{1,7} = -e \quad \bm{\mathcal{A}}_{1,9} = -e\\
\bm{\mathcal{A}}_{2,1} = \kappa\quad
\bm{\mathcal{A}}_{2,2} = -\beta - \mu\\
\bm{\mathcal{A}}_{2,7} = e \quad \bm{\mathcal{A}}_{2,8} = -i \quad \bm{\mathcal{A}}_{2,10} = -i\\
\bm{\mathcal{A}}_{3,1} = \rho\quad
\bm{\mathcal{A}}_{3,2} = \beta\quad
\bm{\mathcal{A}}_{3,3} = -\gamma\\
\bm{\mathcal{A}}_{3,8} = i\quad
\bm{\mathcal{A}}_{3,9} = e\quad
\bm{\mathcal{A}}_{3,11} = -r\\
\bm{\mathcal{A}}_{4,2} = \mu \quad \bm{\mathcal{A}}_{4,10} = i
\end{array}
\label{eq:jacobian}
\end{equation}

\subsection{The extended Kalman filter algorithm for simultaneous parameter and variable tracking\label{sec:EKF}}
Official epidemic data are commonly reported on a daily or weekly bases. Therefore, the measurements used for model variable and parameter estimation are discrete time, while the compartmental model proposed throughout this work is continuous. As a result two approaches are available for implementing an EKF: 1) to use continuous-discrete Kalman filters, which mix continuous state equations with discrete measurement equations, or 2) to discretize the dynamic model of the system and to implement a discrete EKF. The formulation for the latter approach is detailed in Algorithm \ref{alg:EKF}. 
\begin{algorithm}[hbt]
\begin{flushleft}
\caption{An extended Kalman filter for simultaneous compartment variable and model parameter tracking
\label{alg:EKF}}
\begin{algorithmic}[1]
\REQUIRE{Noisy measurements (regular reports) of the epidemic spread $\bm{y}_k$}
\REQUIRE Initial conditions: $\mathbf{Q}$, $\mathbf{R}$, $\hat{\bm{x}}_{0}^+$, $\mathbf{P}_{0}^+$
\ENSURE{$\hat{\bm{x}}_{k}^+$ (vector of state and model parameter estimates)}
\FOR{$k=0\cdots T$}
\STATE \textit{State prediction:}

\STATE $\hat{\bm{x}}_{k+1}^- = \mathbf{f}(\hat{\bm{x}}_{k}^+, \bar{\bm{w}} ; \hat{\bm{\theta}}_{k}^+,k\Delta)$ 
\STATE $\mathbf{P}_{k+1}^- = \bm{\mathcal{A}}_{k}^+\mathbf{P}_{k}^+\bm{\mathcal{A}}_{k}^{+T} + \mathbf{Q}$

\textit{Measurement update:}
\STATE $\mathbf{K}_{k} = \mathbf{P}_{k}^- \mathbf{C}_{k}^{-T}[\mathbf{C}_{k}^-\mathbf{P}_{k}^-\mathbf{C}_{k}^{-T} + \mathbf{R}]^{-1}$

\STATE $\hat{\bm{y}}_{k}^- = \mathbf{g}(\hat{\bm{x}}_{k}^-; \hat{\bm{\theta}}_{k}^-, k\Delta)$

\STATE $\bm{i}_{k} = \bm{y}_{k} - \hat{\bm{y}}_{k}^-$
\STATE $\hat{\bm{\sigma}}_{k}^+ = \hat{\bm{\sigma}}_{k}^- + \mathbf{K}_{k}\bm{i}_k$
\STATE $\mathbf{P}_{k}^+ = [\mathbf{I} - \mathbf{K}_{k}\mathbf{C}_{k}^-]\mathbf{P}_{k}^-$
\STATE \textit{Check and enforce variable and parameter ranges using hard-constraints}
\STATE \textit{Performance monitoring}
\ENDFOR
\end{algorithmic}
\end{flushleft}
\end{algorithm}

\subsection{The extended Kalman filter algorithm for trend tracking from noisy data\label{sec:EKFtrendtracking}}
For pandemic trend smoothing and forecasting described in Section \ref{sec:trendestimation}, we define $y_k = y(k\Delta)$, $\bm{s}_k \stackrel{\Delta}{=} [x(k\Delta), \lambda(k\Delta)]^T$ and $\bm{w}_k \stackrel{\Delta}{=} [w(k\Delta), v(k\Delta)]^T$ as the discrete-time observation, state vector and process noise vector. Accordingly, (\ref{eq:overallmodel}) can be formulated in the discrete form as follows:
\begin{equation}
    \begin{array}{l}
    \bm{s}_{k+1} = \mathbf{f}(\bm{s}_{k}, \bm{w}_k ; h(\mathbf{p}(t)))\\
    y_k = \mathbf{c}^T\bm{s}_{k} + n_k = [1 \quad 0]\bm{s}_{k} + n_k
    \end{array}
\end{equation}
The observation equation is already linear, and the Jacobian of the state equation with respect to the state variables is 
\begin{equation}
\bm{\mathcal{A}}_k \displaystyle\stackrel{\Delta}{=} \frac{\partial \mathbf{f}(\cdot)}{\partial \bm{s}_k} = 
\begin{bmatrix}
\exp(\Delta \lambda_k) & \Delta s_k \exp(\Delta \lambda_k)\\ 0 & \displaystyle\alpha (1 - \tanh(\cdot)^2)
\end{bmatrix}
\end{equation}
where $\tanh(\cdot)$ has the same arguments as in (\ref{eq:overallmodel}). With these definitions, the proposed EKF algorithm is summarized in Algorithm \ref{alg:EKF}. The algorithm can be proved to be asymptotically stable, which implies that even though the new cases might follow an exponential growth, the EKF can track the state vector.

The source codes for this implementation and the training EKS implementation are available in our online repository.
\begin{algorithm}[hbt]
\begin{flushleft}
\caption{An extended Kalman filter for smoothing and forecasting the pandemic trend and its exponential rate
\label{alg:EKF2}}
\begin{algorithmic}[1]
\REQUIRE{Daily reports of the epidemic spread $y_k$}
\REQUIRE Process noise covariance matrix $\mathbf{Q}$, observation noise variance $r$, initial states $\hat{\bm{s}}_{0}^+$ with covariance matrix $\mathbf{P}_{0}^+$, and process noise mean $\bar{\bm{w}}$
\ENSURE{$\hat{\bm{s}}_{k}^+ = [\hat{x}_k , \hat{\lambda}_k]$ (vector of daily state estimates)}
\FOR{$k=0\cdots T$}
\STATE \textit{State prediction:}

\STATE $\hat{\bm{s}}_{k+1}^- = \mathbf{f}(\hat{\bm{s}}_{k}^+, \bar{\bm{w}} ; h(\mathbf{p}_k))$ 
\STATE $\mathbf{P}_{k+1}^- = \bm{\mathcal{A}}_{k}^+\mathbf{P}_{k}^+\bm{\mathcal{A}}_{k}^{+T} + \mathbf{Q}$

\textit{Measurement update:}
\STATE $\mathbf{K}_{k} = \mathbf{P}_{k}^- \mathbf{c}[\mathbf{c}^T\mathbf{P}_{k}^-\mathbf{c} + r]^{-1}$
\STATE $\hat{y}_{k}^- = \mathbf{c}^T\hat{\bm{x}}_{k}^-$
\STATE $i_{k} = y_{k} - \hat{y}_{k}^-$
\STATE $\hat{\bm{s}}_{k}^+ = \hat{\bm{s}}_{k}^- + \mathbf{K}_{k}\bm{i}_k$
\STATE $\mathbf{P}_{k}^+ = [\mathbf{I} - \mathbf{K}_{k}\mathbf{c}^T]\mathbf{P}_{k}^-$
\ENDFOR
\end{algorithmic}
\end{flushleft}
\end{algorithm}
\section*{Acknowledgment}
The first versions of this manuscript were drafted in March 2020, in the midst of the COVID-19 coronavirus outbreak, during the affiliation of R. Sameni with GIPSA-lab, Universit\'e Grenoble Alpes, CNRS, Grenoble INP, Grenoble, France. The author would like to sincerely thank Professor Christian Jutten, Emeritus Professor of Universit\'e Grenoble Alpes, for his insightful and motivating comments throughout this study.
\bibliographystyle{IEEEtran}
\bibliography{IEEEabrv,References}
\end{document}